\documentclass[proceedings]{JHEP3}
\usepackage{amsfonts}
\usepackage{amsmath}
\usepackage{epsfig,multicol}

\newbox\mybox

\newcommand\fverb{\setbox\mybox=\hbox\bgroup\verb}
\newcommand\fverbdo{\egroup\medskip\noindent\fbox{\unhbox\mybox}\ }
\newcommand\fverbit{\egroup\item[\fbox{\unhbox\mybox}]}
\conference{Affine Toda field theories related to 
non-crystallographic Coxeter groups}
\abstract{We propose affine Toda field theories related to the non-crystallographic Coxeter groups
$H_2, H_3$ and $H_4$. The classical mass spectrum, the classical three-point couplings and the 
one-loop corrections to the mass renormalisation are determined. The construction is carried out by means of  
a reduction procedure from crystallographic to non-crystallographic Coxeter groups. The embedding structure
explains for various affine Toda field theories that their particles can be organised in pairs, such that their
relative masses differ by the golden ratio.}

\title{Affine Toda field theories related to Coxeter groups of
non-crystallographic type}
\author{Andreas Fring and Christian Korff \\
Centre for Mathematical Science, City University, \\
Northampton Square, London EC1V 0HB, UK\\
E-mail: \email{A.Fring@city.ac.uk , C.Korff@city.ac.uk}}

\input{tcilatex}

\begin{document}

\section{Introduction}

The Ising model is generally considered as the prime example of integrable
models. When viewed in the continuous limit as a $c=1/2$ conformal field
theory \cite{BPZ}, it is a well known fact that it can be realized as an $%
E_{8}^{(1)}\otimes E_{8}^{(1)}/E_{8}^{(2)}$-coset model \cite{GKO}. Even
when the conformal symmetry is broken, by perturbing the theory with a
primary field of scaling dimension ($1/16,1/16$) \cite{Zamo}, the $E_{8}$
structure survives in form of a (minimal) $E_{8}$-affine Toda field theory
(ATFT) \cite{HM,EY}. It will be one of the results in this paper to show
that there is an even more fundamental structure than $E_{8}$ underlying
this particular model, the non-crystallographic Coxeter group $H_{4}$. We
draw here on the observation made first by Sherbak in 1988 \cite{Scher},
namely that $H_{4}$ can be embedded into $E_{8}$, see also \cite{MP,KKA} for
further developments of these mathematical structures. Loosely speaking, one
may regard the $E_{8}$-theory as two copies of $H_{4}$-theories. We get a
first glimpse of this structure from a more physical point of view when we
bring the mass spectrum of minimal $E_{8}$-affine Toda field theory found
originally in \cite{Zamo} into the form 
\begin{equation}
\begin{array}{llll}
m_{1}=1,\qquad & m_{2}=2\cos \frac{\pi }{30},\qquad & m_{3}=\sqrt{\sin \frac{%
11\pi }{30}/\sin \frac{\pi }{30}},\qquad & m_{4}=2\phi \cos \frac{7\pi }{30},
\\ 
m_{5}=\phi m_{1}, & m_{6}=\phi m_{2}, & m_{7}=\phi m_{3}, & m_{8}=\phi m_{4}.%
\end{array}
\label{m8}
\end{equation}
We have set here the overall mass scale to one. Remarkably, these mass
ratios are the same in the classical as well as in the quantum theory, as
all masses renormalize with an overall factor, see \cite{BCDS2,BCDS,DDV} and
references therein. We observe here that there are four \textquotedblleft
fundamental\textquotedblright\ masses present in the theory, whereas the
other ones can be obtained simply by a multiplication with the golden ratio 
\begin{equation}
\phi =\frac{1}{2}(1+\sqrt{5})=\phi ^{2}-1~.  \label{tau}
\end{equation}
It will turn out that each of the sets ($m_{1},m_{2},m_{3},m_{4}$) and ($%
m_{5},m_{6},m_{7},m_{8}$) can be associated with an $H_{4}$-ATFT.

The other popular integrable quantum field theory is the sine-Gordon model,
see e.g. \cite{ZZ,BFKZ} and references therein. It is a well known fact \cite%
{KF,KKF} that once the coupling constant $\nu $ is taken to be $1/\nu =n$,
with $n$ being an integer, the backscattering amplitudes vanish and the
theory reduces to a minimal $D_{n+1}$-ATFT. In particular for $n=5$ we find
a similar pattern for the mass ratios as discussed above. The $D_{6}$-ATFT
mass spectrum reads up to an overall mass scale 
\begin{equation}
\begin{array}{lll}
m_{1}=\phi ^{-1},\qquad & m_{2}=\sqrt{1+\phi ^{-2}},\qquad & m_{3}=1,\qquad
\\ 
m_{4}=\phi m_{1}, & m_{5}=\phi m_{2}, & m_{6}=\phi m_{3}.%
\end{array}
\label{mD}
\end{equation}
It this case the sets ($m_{1},m_{2},m_{3}$) and ($m_{4},m_{5},m_{6}$) can be
associated with an $H_{3}$-ATFT.

The above mentioned structure can be explained simply by the fact that $%
H_{4} $ can be embedded into $E_{8}$ and $H_{3}$ into $D_{6}$, such that the
non-crystallographic structure is \textquotedblleft
visible\textquotedblright\ inside the theories related to crystallographic
Coxeter groups.

Besides having a \textquotedblleft non-crystallographic
pattern\textquotedblright\ inside theories related to crystallographic
Coxeter groups, it is interesting to ask the question whether it is possible
to construct theories purely based on these latter groups. In particular, $%
H_{3}$ being a three-dimensional symmetry group of the icosahedron, a
regular solid with 20 triangle faces, finds a natural application in
physical \cite{HP,Bak,FYL}, chemical \cite{MKLE,CF} an even biological
systems \cite{Cas,V1,R2,V2,V3}. In the context of integrable (solvable)
models, Calogero-Sutherland models have been formulated based also on $H_{3}$
\cite{HR,R1} and it should be possible to extend these investigations to
other non-crystallographic Coxeter groups. However, so far no ATFT for such
type of group has been considered, the main reason being that unlike for
crystallographic ones, in this case there is no Lie algebra at disposal,
which is vital in that context. This deficiency can be overcome by
exploiting the embedding structure and reduce the theories associated to
crystallographic Coxeter groups to new types of theories related entirely to
non-crystallographic ones. This is somewhat similar in spirit to the folding
procedure carried out by Olive and Turok \cite{DIO}, who constructed ATFT
for non-simply laced Lie algebras from those related to simply laced
algebras, exploiting the embedding of the former into the latter. However,
in comparison, there is one crucial difference. Whereas in the folding
scenario \cite{DIO} the reduced models are identical to a formulation purely
in terms of non-simply laced algebras, the models we obtain here vitally
rely in their construction on the embedding and can not be formulated
directly in terms of non-crystallographic Coxeter groups on the level of the
Lagrangian.

Our manuscript is organized as follows: In section 2 we review and develop
the mathematics associated to the embedding of non-crystallographic into
crystallographic Coxeter groups. In section 3 we apply these notions to
affine Toda field theory and construct in particular their classical mass
spectra and fusing structures. In addition we start the development of a
quantum field theory by computing the mass renormalisations. We state our
conclusions in section 4. In the appendix we present explicit computations
of various orbits of coloured simple roots related to non-crystallographic
and crystallographic Coxeter groups and exhibit how they can be embedded
into one another.

\section{Embedding of non-crystallographic into crystallographic Coxeter
groups}

Coxeter graphs are finite graphs, whose edges are labelled by some integers $%
m_{ij}$ joining the vertices $i$ and $j$ \cite{Hum2}. To each of these
graphs one can associate a finite reflection group. When the
crystallographic condition is satisfied, that is $m_{ij}$ for $i$ $\neq j$
takes only the values $2,3,4$ or $6$, these groups are Weyl groups. In
contrast to the non-crystallographic groups, the crystallographic ones can
be related to Lie algebras and Lie groups. Lie theory is exploited largely
in the context of integrable models, which is one of the main reasons why
crystallographic groups enjoy wider applications. Here we use the embedding
of non-crystallographic groups into crystallographic ones, such that we can
still exploit the Lie structure related to the larger groups.

\subsection{Root systems}

In order to assemble the necessary mathematical notions, we start by
introducing a map $\omega $ from a root system$~\Delta ^{c}$ which is
invariant under the action of a crystallographic Coxeter group $\mathcal{W}$
of rank $\ell $ into the union of two sets $\tilde{\Delta}^{nc}$ related to
a non-crystallographic group $\mathcal{\tilde{W}}$ of rank $\tilde{\ell}%
=\ell /2$ 
\begin{equation}
\omega :~\Delta ^{c}\rightarrow \tilde{\Delta}^{nc}\cup \phi \tilde{\Delta}%
^{nc}~.  \label{pi}
\end{equation}%
Throughout this manuscript we adopt the notation that quantities related to
crystallographic and non-crystallographic groups are specified with the same
symbol, distinguished by an additional tilde, e.g. $\alpha \in \Delta
^{c}=\Delta $ and $\tilde{\alpha}\in \tilde{\Delta}^{nc}=\tilde{\Delta}$.
Introducing a special labelling for the vertices on the Coxeter graphs, or
equivalently the simple roots, we can always realize this map as 
\begin{equation}
\alpha _{i}\mapsto \omega (\alpha _{i})=\left\{ 
\begin{array}{ll}
\tilde{\alpha}_{i} & \text{for\quad }1\leq i\leq \tilde{\ell}=\ell /2 \\ 
\phi \tilde{\alpha}_{i-\tilde{\ell}}\qquad & \text{for\quad }\tilde{\ell}%
<i\leq \ell .%
\end{array}%
\right.  \label{pi2}
\end{equation}%
Our labelling allows for a generic treatment of the embedding and differs
for instance from the one used in \cite{Scher,MP,KKA}. A further important
property guaranteed by our conventions is $\alpha _{i}\cdot \alpha _{i+%
\tilde{\ell}}=0$. Both types of root systems $\Delta $ and $\tilde{\Delta}$
are equipped with a symmetric bilinear form or inner product. In \cite{MP}
Moody and Patera noticed the remarkable fact that the map $\omega $ is an
isometric isomorphism, such that we may compute inner products in the root
system $\Delta $ from inner products in $\tilde{\Delta}$ 
\begin{equation}
\alpha \cdot \beta =R(\omega (\alpha )\cdot \omega (\beta )).  \label{RF}
\end{equation}%
Here the map $R$, called a rational form relative to $\phi $, extracts from
a number of the form\footnote{%
Note that higher powers of $\phi $ can be reduced to that form by a repeated
use of (\ref{tau}), e.g. $\phi ^{2}=1+\phi $, $\phi ^{3}=1+2\phi $, $\phi
^{4}=2+3\phi $, $\phi ^{5}=3+5\phi $,$\ldots $, $\phi ^{n}=f_{n-1}+\phi
f_{n} $ where $f_{n}$ is the $n$-th Fibonacci number obeying the recursive
relation $f_{n+1}=$ $f_{n}+$ $f_{n-1}$.} $a+\phi b$ with $a,b\in \mathbb{Q}$
the rational part $a$ 
\begin{equation}
R(a+\phi b)=a.
\end{equation}%
We normalize all our roots to have length 2, such that $\alpha ^{2}=$ $%
\tilde{\alpha}^{2}=2$ for $\alpha \in \Delta $ and $\tilde{\alpha}\in \tilde{%
\Delta}$. According to (\ref{RF}) we may therefore compute the Cartan matrix
related to $\Delta $ entirely from inner products in $\tilde{\Delta}$ 
\begin{equation}
K_{ij}=\frac{2\alpha _{i}\cdot \alpha _{j}}{\alpha _{j}\cdot \alpha _{j}}%
=R(\omega (\alpha _{i})\cdot \omega (\alpha _{j}))~=R(\phi ^{t_{i}+t_{j}}%
\tilde{\alpha}_{i-t_{i}\tilde{\ell}}\cdot \tilde{\alpha}_{j-t_{j}\tilde{\ell}%
}).  \label{KN}
\end{equation}%
where $t_{i}=0$ for $1\leq i\leq \tilde{\ell}$ and $t_{i}=1$ for $\tilde{\ell%
}<i\leq \ell $. For our purposes it will be most important to achieve also
the opposite, which can not be found in \cite{MP}, namely to compute inner
products in $\tilde{\Delta}$ from those in $\Delta $. For this aim we
introduce here the map 
\begin{equation}
\tilde{\omega}:~\tilde{\Delta}\rightarrow \Delta \oplus \phi \Delta ,~
\end{equation}%
which acts on the simple roots in $\tilde{\Delta}$ as 
\begin{equation}
\tilde{\alpha}_{i}\mapsto \tilde{\omega}(\tilde{\alpha}_{i})=(\alpha
_{i}+\phi \alpha _{i+\tilde{\ell}})\qquad \text{for~~}1\leq i\leq \tilde{\ell%
}~.  \label{at}
\end{equation}%
Note that $\tilde{\omega}(\tilde{\alpha}_{i})^{2}=2+2\phi ^{2}$, such that $%
\tilde{\omega}$ is not an isometry. Instead, we find that inner products in $%
\tilde{\Delta}$ are related to inner products in $\Delta $ by means of 
\begin{equation}
\omega (\alpha _{i})\cdot \tilde{\alpha}_{j}=\alpha _{i}\cdot \tilde{\omega}(%
\tilde{\alpha}_{j})\qquad \text{for~~}1\leq j\leq \tilde{\ell}~,1\leq i\leq
\ell ~.  \label{oot}
\end{equation}%
Expanding (\ref{oot}) yields immediately a relation between the Cartan
matrices of $\Delta $ and $\tilde{\Delta}$%
\begin{equation}
K_{ij}+\phi K_{i(j+\tilde{\ell})}=\left\{ 
\begin{array}{ll}
\tilde{K}_{ij} & \text{for~~}1\leq i\leq \tilde{\ell}~ \\ 
\phi \tilde{K}_{(i-\tilde{\ell})j}\qquad & \text{for~~}\tilde{\ell}~<i\leq
\ell .~%
\end{array}%
\right.  \label{KK}
\end{equation}%
Noting that for the Coxeter groups we consider the Cartan matrices are
symmetric, such that the relations (\ref{KK}) also hold with $%
i\leftrightarrow j$. As the right hand side involves only one inner product
in $\tilde{\Delta}$, this formula achieves our objective and we may now
express inner products in $\tilde{\Delta}$ in terms of those in $\Delta $.
Recalling that simple roots and fundamental weights $\lambda _{i}$ are
related as $\alpha _{i}=\sum_{j}K_{ij}\lambda _{j}$ and $\tilde{\alpha}%
_{i}=\sum_{j}\tilde{K}_{ij}\tilde{\lambda}_{j}$, it follows directly that (%
\ref{oot}) also holds when we replace simple roots by fundamental weights.
As $K_{ij}^{-1}=\lambda _{i}\cdot \lambda _{j}$, this means that (\ref{KK})
also holds for the inverse matrices 
\begin{equation}
K_{ij}^{-1}+\phi K_{i(j+\tilde{\ell})}^{-1}=\left\{ 
\begin{array}{ll}
\tilde{K}_{ij}^{-1} & \text{for~~}1\leq i\leq \tilde{\ell}~ \\ 
\phi \tilde{K}_{(i-\tilde{\ell})j}^{-1}\qquad & \text{for~~}\tilde{\ell}%
~<i\leq \ell ~%
\end{array}%
\right.
\end{equation}%
and (\ref{pi2}), (\ref{at}) for $\alpha _{i}\rightarrow \lambda _{i}$, $%
\tilde{\alpha}_{i}\rightarrow $ $\tilde{\lambda}_{i}$. With regard to our
application it will be particularly important to relate the eigensystems $K$
and $\tilde{K}$. Abbreviating $\kappa _{ij}=K_{ij}$ and $\hat{\kappa}%
_{ij}=K_{i(j+\tilde{\ell})}$ for $1\leq i,j\leq \tilde{\ell}$ we can
block-decompose the Cartan matrix $K$ further and verify that 
\begin{equation}
U^{-1}KU=U^{-1}\left( 
\begin{array}{cc}
\kappa & \hat{\kappa} \\ 
\hat{\kappa} & \kappa +\hat{\kappa}%
\end{array}%
\right) U=\left( 
\begin{array}{cc}
\tilde{K}(\phi ) & 0 \\ 
0 & \tilde{K}(-\phi ^{-1})%
\end{array}%
\right) ~~~\text{with~~}U=\left( 
\begin{array}{rr}
\mathbb{I} & \mathbb{I} \\ 
\phi \mathbb{I} & -\mathbb{I}/\phi%
\end{array}%
\right) ,  \label{sim}
\end{equation}%
where $\mathbb{I}$ is the $\tilde{\ell}\times \tilde{\ell}$ unit matrix.
This means of course that the $\ell $ eigenvalues of $K$, which are known to
be of the form $e_{n}=4\sin ^{2}(\pi s_{n}/2h)$ with $h$ being the Coxeter
number (see below for more details). The $s_{n}$ label the $\ell $ exponents
of $\mathcal{W}$ and are identical to the union of the $\tilde{\ell}~$%
eigenvalues of $\tilde{K}(\phi )$ and $\tilde{K}(-\phi ^{-1})$ 
\begin{equation}
\mathcal{S}=\{s_{1},s_{2},\ldots ,s_{\ell }\}=\mathcal{\tilde{S}(}\phi 
\mathcal{)}\cup \mathcal{\tilde{S}}^{\prime }\mathcal{(-}\phi ^{-1}\mathcal{)%
}.  \label{SSS}
\end{equation}%
The eigenvalues are invariant under the change $\phi \rightarrow -\phi $.
Labelling the $\ell $ eigenvectors of $K$ by $y_{n}$, such that $%
Ky_{n}=e_{n}y_{n}$, we can construct the eigenvectors $\tilde{y}_{\tilde{s}%
}(\phi )$ and $\tilde{y}_{\tilde{s}^{\prime }}(-\phi ^{-1})$ of $\tilde{K}%
(\phi )$ and $\tilde{K}(-\phi ^{-1})$, respectively, from (\ref{sim}) as 
\begin{equation}
U^{-1}y_{s}=\tilde{y}_{\tilde{s}}(\phi )\otimes \tilde{y}_{\tilde{s}^{\prime
}}(-\phi ^{-1}).  \label{v}
\end{equation}%
Conversely, we can construct the eigenvectors of $K$ from the knowledge of
the eigenvectors of $\tilde{K}(\phi )$ and $\tilde{K}(-\phi ^{-1})$ 
\begin{equation}
y_{\tilde{s}}=\tilde{y}_{\tilde{s}}(\phi )\otimes \phi \tilde{y}_{\tilde{s}%
}(\phi )~\qquad \text{and\qquad }y_{\tilde{s}^{\prime }}=\tilde{y}_{\tilde{s}%
^{\prime }}(-\phi ^{-1})\otimes (-\phi ^{-1})\tilde{y}_{\tilde{s}^{\prime
}}(-\phi ^{-1})~.  \label{yn}
\end{equation}%
The first identity follows by exploiting (\ref{KK}) 
\begin{equation}
Ky_{\tilde{s}}=K\binom{\tilde{y}_{\tilde{s}}(\phi )}{\phi \tilde{y}_{\tilde{s%
}}(\phi )}=\left( 
\begin{array}{cc}
\kappa & \hat{\kappa} \\ 
\hat{\kappa} & \kappa +\hat{\kappa}%
\end{array}%
\right) \binom{\tilde{y}_{\tilde{s}}(\phi )}{\phi \tilde{y}_{\tilde{s}}(\phi
)}=\left( 
\begin{array}{cc}
\tilde{K}(\phi ) & 0 \\ 
0 & \tilde{K}(\phi )%
\end{array}%
\right) \binom{\tilde{y}_{\tilde{s}}(\phi )}{\phi \tilde{y}_{\tilde{s}}(\phi
)}.
\end{equation}%
The second relation in (\ref{yn}) is obtained by the same argumentation with 
$\phi \rightarrow -\phi ^{-1}$. These facts will not only be crucial to
formulate new types of ATFT, but also to explain patterns in well studied
models. The knowledge of the distribution of the exponents with respect to
the embedding is important as they grade the conserved charges. The
classical masses are known \cite{BCDS,Mass1,Mass2} to organise as components
of the Perron-Frobenius vector $y_{1}$, such that (\ref{yn}) explains the
aforementioned mass patterns (\ref{m8}) and (\ref{mD}).

\subsection{Coxeter groups}

Having related the root systems $\tilde{\Delta}$ and $\Delta $, let us see
next how to relate the corresponding Coxeter groups $\mathcal{W}$ and $%
\mathcal{\tilde{W}}$, which leave them invariant. We recall \cite{Hum,Hum2}
that Coxeter groups are generated by reflections on the hyperplane through
the origin orthogonal to simple roots $\alpha _{i}$ 
\begin{equation}
\sigma _{i}(x)=x-2\frac{x\cdot \alpha _{i}}{\alpha _{i}\cdot \alpha _{i}}%
\alpha _{i}~\ \qquad \text{for }1\leq i\leq \ell ,~x\in \mathbb{R}^{\ell }.
\label{W}
\end{equation}
We can then think of the Coxeter group $\mathcal{W}$ as the set of all words
in the generators $\{\sigma _{i}\}$ subject to the relations 
\begin{equation}
~(\sigma _{i}\sigma _{j})^{m_{ij}}=1,\qquad ~1\leq i,j\leq \ell ,  \label{WW}
\end{equation}
where the $m_{ij}$ can take the values $1,2,3,4,5$ or $6$. Here we focus on
the Coxeter groups for which the Cartan matrix is symmetric, in which case $%
m_{ij}=\pi \arccos ^{-1}(-K_{ij}/2)$ are the integers labelling the edges of
the Coxeter graph mentioned at the beginning of this section. $\mathcal{%
\tilde{W}}$ is constructed analogously when replacing $K$ by $\tilde{K}(\phi
)$. Note that when we use instead of $\tilde{K}(\phi )$ the matrix $\tilde{K}%
(-\phi ^{-1})$, which naturally emerges through (\ref{sim}), this will lead
to the same Coxeter relations.

From a group theoretical point of view we can identify $\tilde{\sigma}%
_{i}\hookrightarrow \sigma _{i}\sigma _{i+\tilde{\ell}}$. With the help of
the map $\omega $ we can relate the reflections $\tilde{\sigma}_{i}$
building up $\mathcal{\tilde{W}}$ to those constituting $\mathcal{W}$ as 
\begin{equation}
\tilde{\sigma}_{i}\omega =\omega \sigma _{i}\sigma _{i+\tilde{\ell}}~\qquad
\qquad \text{for~~}1\leq i\leq \tilde{\ell}~.  \label{ss}
\end{equation}
This is seen easily by acting on a simple root in $\Delta $ 
\begin{eqnarray}
\omega \sigma _{i}\sigma _{i+\tilde{\ell}}(\alpha _{j}) &=&\omega \left[
\alpha _{j}-(\alpha _{j}\cdot \alpha _{i+\tilde{\ell}})\alpha _{i+\tilde{\ell%
}}-(\alpha _{j}\cdot \alpha _{i})\alpha _{i}\right]  \label{12} \\
&=&\omega (\alpha _{j})-\phi (\alpha _{j}\cdot \alpha _{i+\tilde{\ell}})%
\tilde{\alpha}_{i}-(\alpha _{j}\cdot \alpha _{i})\tilde{\alpha}_{i},
\label{13}
\end{eqnarray}
where in (\ref{12}) we used twice (\ref{W}) and the fact that in our
labelling we always have $\alpha _{i}\cdot \alpha _{i+\tilde{\ell}}=0$. Then
(\ref{13}) simply follows upon using (\ref{pi2}). On the other hand using (%
\ref{W}) for $\mathcal{\tilde{W}}$ and (\ref{oot}) thereafter, we obtain 
\begin{eqnarray}
\tilde{\sigma}_{i}\omega (\alpha _{j}) &=&\omega (\alpha _{j})-(\omega
(\alpha _{j})\cdot \tilde{\alpha}_{i})\tilde{\alpha}_{i} \\
&=&\omega (\alpha _{j})-\phi (\alpha _{j}\cdot \alpha _{i+\tilde{\ell}})%
\tilde{\alpha}_{i}-(\alpha _{j}\cdot \alpha _{i})\tilde{\alpha}_{i}.
\end{eqnarray}

\noindent In \cite{MP} a similar identification has been made for the
specific case of the embedding $H_{4}\hookrightarrow E_{8}$, which relies on
the property of an inflation map which mimics the action of $\omega $
entirely inside $\Delta $. Here we avoid the introduction of such a quantity.

\noindent Furthermore, for the second map $\tilde{\omega}$ we have the
supplementary identity 
\begin{equation}
\tilde{\omega}\tilde{\sigma}_{i}=\sigma _{i}\sigma _{i+\tilde{\ell}}~\tilde{%
\omega}\qquad \qquad \text{for~~}1\leq i\leq \tilde{\ell}~,  \label{23}
\end{equation}%
which follows from a similar argument as (\ref{ss}) upon using (\ref{W}), (%
\ref{at}) and (\ref{KK}). As we saw already in (\ref{oot}), we note here
that $\tilde{\omega}$ plays the role of the \textquotedblleft
inverse\textquotedblright\ of $\omega $. There is no analogue to the
inflation map in this case. More precisely we find 
\begin{equation}
\omega \tilde{\omega}=(1+\phi ^{2})\;\mathbb{I\qquad }\text{and\qquad }%
\tilde{\omega}\omega =\left( 
\begin{array}{ll}
\mathbb{I} & \phi \mathbb{I} \\ 
\phi \mathbb{I} & \phi ^{2}\mathbb{I}%
\end{array}%
\right) ,
\end{equation}%
where $\mathbb{I}$ is the $\tilde{\ell}\times \tilde{\ell}$ unit matrix
already encountered in (\ref{sim}).

For the application we have in mind, it is important to note that the entire
root system $\Delta $ can be separated into orbits $\Omega _{i}$, such that $%
\Delta =\bigcup\nolimits_{i=1}^{\ell }\Omega _{i}$, each containing $h$
roots. Here $h$ is the order of the Coxeter element $\sigma $, i.e. the
Coxeter number already introduced after (\ref{sim}), which is a product of $%
\ell $ simple reflections with the property $\sigma ^{h}=1$. As the
reflections do not commute in general, such that Coxeter elements only form
conjugacy classes, we have to specify our conventions. For this purpose we
attach values $c_{i}=\pm 1$ to the vertices $i$ of the Coxeter graph, in
such a way that no two vertices related to the same value are linked
together. The vertices then separate into two disjoint sets $V_{\pm }$ and
the Coxeter element $\sigma =\prod\nolimits_{i\in V_{-}}\sigma _{i}$ $%
\prod\nolimits_{i\in V_{+}}\sigma _{i}$ is uniquely defined. Introducing
``coloured''\ simple roots as $\gamma _{i}=c_{i}\alpha _{i}$, each orbit $%
\Omega _{i}(\tilde{\Omega}_{i})$ is then generated by $h(\tilde{h})$
successive actions of $\sigma (\tilde{\sigma})$ on $\gamma _{i}(\tilde{\gamma%
}_{i})$ \cite{PD1,FO}. Since we know how to relate the simple reflections of 
$\mathcal{W}$ and $\mathcal{\tilde{W}}$ by means of (\ref{ss}) and (\ref{23}%
), it is obvious that the Coxeter elements are intertwined as 
\begin{equation}
\tilde{\sigma}\omega =\omega \sigma \qquad \text{and\qquad }\tilde{\omega}%
\tilde{\sigma}=\sigma \tilde{\omega}.  \label{sst}
\end{equation}
Note further that 
\begin{equation}
\tilde{c}_{i}=c_{i}=c_{i+\tilde{\ell}}.  \label{cc}
\end{equation}
which is important for (\ref{sst}) to work, as it guarantees that (\ref{ss})
is not an obstacle for the above mentioned separation of the product into
different colours. We then find that 
\begin{equation}
\omega (\Omega _{i})=\left\{ 
\begin{array}{ll}
\tilde{\Omega}_{i} & \text{for\quad }1\leq i\leq \tilde{\ell}=\ell /2 \\ 
\phi \tilde{\Omega}_{i}\qquad & \text{for\quad }\tilde{\ell}<i\leq \ell .%
\end{array}
\right. ~  \label{PO}
\end{equation}
Thus we can realize the map (\ref{pi}) orbit by orbit.

After this generic preliminaries let us discuss in detail the concrete
examples of the embeddings $H_{2}\hookrightarrow A_{4}$, $%
H_{3}\hookrightarrow D_{6}$ and $H_{4}\hookrightarrow E_{8}$.

\subsection{The embedding $H_{2}\hookrightarrow A_{4}$}

We start with the most simple example, that is the embedding of $H_{2}$
(also referred to as $I(5)$, see e.g. \cite{Hum2}) into $A_{4}$. First of
all we have to fix our conventions for naming the simple roots, which we do
by means of the following Coxeter graph (where we adopt the common rule \cite%
{Hum,Hum2} that the label $m_{ij}=3$ corresponds to one lace)

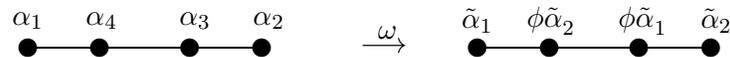
\unitlength=0.680000pt 
\begin{picture}(300.0,70.00)(150.00,125.00)

\put(389.00,165.00){\makebox(0.00,0.00){${\alpha}_{2}$}}
\put(348.0,165.00){\makebox(0.00,0.00){${\alpha}_{3}$}}
\put(296.00,165.00){\makebox(0.00,0.00){${\alpha}_4$}}
\put(255.00,165.00){\makebox(0.00,0.00){${\alpha}_1$}}

\put(350.00,150.00){\line(1,0){30.00}}
\put(330.00,150.00){\line(1,0){10.00}}
\put(300.00,150.00){\line(1,0){30.00}}
\put(260.00,150.00){\line(1,0){30.00}}

\put(385.00,150.00){\circle*{10.00}}
\put(345.00,150.00){\circle*{10.00}}
\put(255.00,150.00){\circle*{10.00}}
\put(295.00,150.00){\circle*{10.00}}
\put(450.00,155.00){$\omega$}
\put(440.00,147.00){$ \longrightarrow $}

\put(638.00,165.00){\makebox(0.00,0.00){${\tilde{\alpha}}_2$}}
\put(597.00,165.00){\makebox(0.00,0.00){${\phi \tilde{\alpha}}_1$}}

\put(546.00,165.00){\makebox(0.00,0.00){${\phi \tilde{\alpha}}_2$}}
\put(505.00,165.00){\makebox(0.00,0.00){${ \tilde{ \alpha}}_1$}}

\put(600.00,150.00){\line(1,0){30.00}}
\put(580.00,150.00){\line(1,0){10.00}}
\put(550.00,150.00){\line(1,0){30.00}}
\put(510.00,150.00){\line(1,0){30.00}}

\put(635.00,150.00){\circle*{10.00}}
\put(595.00,150.00){\circle*{10.00}}
\put(505.00,150.00){\circle*{10.00}}
\put(545.00,150.00){\circle*{10.00}}
\end{picture}

\noindent These conventions guarantee that we can realize the map $\omega $
as defined in (\ref{pi2}), which is also indicated in the above diagrams.
Accordingly, the Cartan matrix of $A_{4}$, as defined in general in (\ref{KN}%
), reads 
\begin{equation}
K=\left( 
\begin{array}{rrrr}
2 & 0 & 0 & -1 \\ 
0 & 2 & -1 & 0 \\ 
0 & -1 & 2 & -1 \\ 
-1 & 0 & -1 & 2%
\end{array}
\right) =R\left( 
\begin{array}{rr}
\tilde{K} & \phi \tilde{K} \\ 
\phi \tilde{K} & \phi ^{2}\tilde{K}%
\end{array}
\right) ,  \label{6}
\end{equation}
where the Cartan matrix of $H_{2}$ is 
\begin{equation}
\tilde{K}=\left( 
\begin{array}{rr}
2 & -\phi \\ 
-\phi & 2%
\end{array}
\right) .  \label{7}
\end{equation}
We may read off relation (\ref{KK}) directly by comparing (\ref{6}) with (%
\ref{7}). Furthermore, we note from the second relation in (\ref{6}) that
the identity (\ref{KN}) holds by reducing the higher powers of $\phi $ as
indicated above. The Coxeter numbers are $h=\tilde{h}=5$ and the set of
exponents separate according to (\ref{SSS}) into 
\begin{equation}
\{1,2,3,4\}=\{1,4\}\cup \{2,3\}.
\end{equation}
Let us now see in detail how $\omega $ acts on the orbits $\Omega _{i}$ and
how the map (\ref{PO}) is realized. We choose $\sigma =$ $\sigma _{1}\sigma
_{3}\sigma _{2}\sigma _{4}$ and $\tilde{\sigma}=$ $\tilde{\sigma}_{1}\tilde{%
\sigma}_{2}$ for the Coxeter element of $A_{4}$ and $H_{2}$, respectively.
The corresponding orbits $\Omega _{i}$ and $\tilde{\Omega}_{i}$ are computed
by successive actions of $\sigma $ and $\tilde{\sigma}$, respectively, on
the simple roots. One realizes that the map $\omega $ relates them indeed as
specified in (\ref{PO}). Indicating in the first column the elements $\sigma
^{p}$ $(\tilde{\sigma}^{p})$ for $1\leq p\leq h=\tilde{h}$ with which we act
on the roots reported in the second row, we find for instance

\begin{center}
\begin{tabular}{||l||c|c|c|c||}
\hline
& $\Omega _{1}$ & $\omega (\Omega _{1})=\tilde{\Omega}_{1}$ & $\Omega _{3}$
& $\omega (\Omega _{3})=\phi \tilde{\Omega}_{1}$ \\ \hline\hline
$\sigma ^{0},\tilde{\sigma}^{0}$ & ${{\alpha }_{1}}$ & ${{\tilde{\alpha}}_{1}%
}$ & ${{\alpha }_{3}}$ & $\phi {{\tilde{\alpha}}_{1}}$ \\ \hline
$\sigma ^{1},\tilde{\sigma}^{1}$ & ${{\alpha }_{3}}+{{\alpha }_{4}}$ & $\phi
({{\tilde{\alpha}}_{1}}+{{\tilde{\alpha}}_{2})}$ & ${{\alpha }_{1}}+{{\alpha 
}_{2}}+{{\alpha }_{3}}+{{\alpha }_{4}}$ & $\phi ^{2}({{\tilde{\alpha}}_{1}}+{%
{\tilde{\alpha}}_{2})}$ \\ \hline
$\sigma ^{2},\tilde{\sigma}^{2}$ & ${{\alpha }_{2}}$ & ${{\tilde{\alpha}}_{2}%
}$ & ${{\alpha }_{4}}$ & $\phi {{\tilde{\alpha}}_{2}}$ \\ \hline
$\sigma ^{3},\tilde{\sigma}^{3}$ & $-{{\alpha }_{2}}-{{\alpha }_{3}}$ & $%
-\phi {{\tilde{\alpha}}_{1}}-{{\tilde{\alpha}}_{2}}$ & $-{{{\alpha }_{1}}}-{{%
{\alpha }_{3}}}-{{{\alpha }_{4}}}$ & $-\phi ^{2}{{\tilde{\alpha}}_{1}}-\phi {%
{\tilde{\alpha}}_{2}}$ \\ \hline
$\sigma ^{4},\tilde{\sigma}^{4}$ & $-{{\alpha }_{1}}-{{\alpha }_{4}}$ & $-{{%
\tilde{\alpha}}_{1}}-\phi {{\tilde{\alpha}}_{2}}$ & $-{{{\alpha }_{2}}}-{{{%
\alpha }_{3}}}-{{{\alpha }_{4}}}$ & $-\phi {{\tilde{\alpha}}_{1}}-\phi ^{2}{{%
\tilde{\alpha}}_{2}}$ \\ \hline
$\sigma ^{5},\tilde{\sigma}^{5}$ & ${{\alpha }_{1}}$ & ${{\tilde{\alpha}}_{1}%
}$ & ${{\alpha }_{3}}$ & $\phi {{\tilde{\alpha}}_{1}}$ \\ \hline
\end{tabular}
\end{center}

\noindent In order to establish the last identification $\omega (\Omega
_{3})=\phi \tilde{\Omega}_{1}$ we simply need to make use of relation (\ref%
{tau}). Furthermore, we obtain for the remaining orbits

\begin{center}
\begin{tabular}{||l||c|c|c|c||}
\hline
& $\Omega _{2}$ & $\omega (\Omega _{2})=\tilde{\Omega}_{2}$ & $\Omega _{4}$
& $\omega (\Omega _{4})=\phi \tilde{\Omega}_{2}$ \\ \hline\hline
$\sigma ^{0},\tilde{\sigma}^{0}$ & $-{{\alpha }_{2}}$ & $-{{\tilde{\alpha}}%
_{2}}$ & $-{{\alpha }_{4}}$ & $-\phi {{\tilde{\alpha}}_{2}}$ \\ \hline
$\sigma ^{1},\tilde{\sigma}^{1}$ & ${{\alpha }_{2}}+{{\alpha }_{3}}$ & $\phi 
{{\tilde{\alpha}}_{1}}+{{\tilde{\alpha}}_{2}}$ & ${{\alpha }_{1}}+{{\alpha }%
_{3}}+{{\alpha }_{4}}$ & $\phi ^{2}{{\tilde{\alpha}}_{1}}+\phi {{\tilde{%
\alpha}}_{2}}$ \\ \hline
$\sigma ^{2},\tilde{\sigma}^{2}$ & ${{\alpha }_{1}}+{{\alpha }_{4}}$ & ${{%
\tilde{\alpha}}_{1}}+\phi {{\tilde{\alpha}}_{2}}$ & ${{\alpha }_{2}}+{{%
\alpha }_{3}}+{{\alpha }_{4}}$ & $\phi {{\tilde{\alpha}}_{1}}+\phi ^{2}{{%
\tilde{\alpha}}_{2}}$ \\ \hline
$\sigma ^{3},\tilde{\sigma}^{3}$ & $-{{\alpha }_{1}}$ & $-{{\tilde{\alpha}}%
_{1}}$ & $-{{{\alpha }_{3}}}$ & $-\phi {{\tilde{\alpha}}_{1}}$ \\ \hline
$\sigma ^{4},\tilde{\sigma}^{4}$ & $-{{\alpha }_{3}}-{{\alpha }_{4}}$ & $%
-\phi ({{\tilde{\alpha}}_{1}}+{{\tilde{\alpha}}_{2})}$ & $-{{\alpha }_{1}}-{{%
\alpha }_{2}}-{{\alpha }_{3}}-{{\alpha }_{4}}$ & $-\phi ^{2}({{\tilde{\alpha}%
}_{1}}+{{\tilde{\alpha}}_{2})}$ \\ \hline
$\sigma ^{5},\tilde{\sigma}^{5}$ & $-{{\alpha }_{2}}$ & $-{{\tilde{\alpha}}%
_{2}}$ & $-{{\alpha }_{4}}$ & $-\phi {{\tilde{\alpha}}_{2}}$ \\ \hline
\end{tabular}
\end{center}

\noindent The identification $\omega (\Omega _{4})=\phi \tilde{\Omega}_{2}$
follows upon using (\ref{tau}). Having established how the root system $%
\Delta $ can be mapped into $\tilde{\Delta}\cup \phi \tilde{\Delta}$ orbit
by orbit, we want to see next how $\mathcal{W}$ relates to $\mathcal{\tilde{W%
}}$. In principle we have to check all $|\mathcal{\tilde{W}}|$ relations in (%
\ref{WW}). However, it suffices to establish the identification (\ref{ss})
for the generating relations of $\mathcal{\tilde{W}}$. It is known that $%
H_{2}$ can be generated entirely from $\tilde{\sigma}_{i}^{2}=1$ for $%
i=1,2,3 $ together with 
\begin{equation}
(\tilde{\sigma}_{1}\tilde{\sigma}_{2})^{5}=1,
\end{equation}
which follows directly from the previous tables, since $\sigma =\tilde{\sigma%
}_{1}\tilde{\sigma}_{2}$. In $A_{4}$ this corresponds to 
\begin{equation}
(\sigma _{1}\sigma _{3}\sigma _{2}\sigma _{4})^{5}=1~.
\end{equation}

\noindent The remaining relations are trivially satisfied. We know that by
definition $\sigma _{i}^{2}=1$ and therefore this also holds when squaring
the product of the embedding on the right hand side of (\ref{ss}) $(\sigma
_{i}\sigma _{i+\tilde{\ell}})^{2}=\sigma _{i}^{2}\sigma _{i+\tilde{\ell}%
}^{2}~=1$. We used here the last equality in (\ref{cc}), such that $\sigma
_{i}$ and $\sigma _{i+\tilde{\ell}}$ commute.

\subsection{The embedding $H_{3}\hookrightarrow D_{6}$}

In this case we fix our conventions as

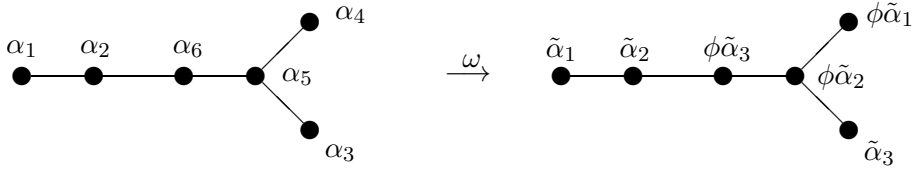
\unitlength=0.680000pt 
\begin{picture}(237.92,150.00)(220.00,65.00)
\put(432.0,107.17){\makebox(0.00,0.00){$\alpha_{3}$}}
\put(437.92,184.17){\makebox(0.00,0.00){${\alpha}_{4}$}}
\put(408.17,149.59){\makebox(0.00,0.00){${\alpha}_{5}$}}
\put(347.50,165.00){\makebox(0.00,0.00){${\alpha}_{6}$}}
\put(296.00,165.00){\makebox(0.00,0.00){${\alpha}_2$}}
\put(255.00,165.00){\makebox(0.00,0.00){${\alpha}_1$}}
\put(389.00,146.33){\line(1,-1){22.67}}
\put(411.67,176.67){\line(-1,-1){23.00}}
\put(350.00,150.00){\line(1,0){30.00}}
\put(330.00,150.00){\line(1,0){10.00}}
\put(300.00,150.00){\line(1,0){30.00}}
\put(260.00,150.00){\line(1,0){30.00}}
\put(415.00,120.00){\circle*{10.00}}
\put(415.00,180.00){\circle*{10.00}}
\put(385.00,150.00){\circle*{10.00}}
\put(345.00,150.00){\circle*{10.00}}
\put(255.00,150.00){\circle*{10.00}}
\put(295.00,150.00){\circle*{10.00}}
\put(500.00,155.00){$\omega$}
\put(490.00,147.00){$ \longrightarrow $}
\put(732.0,107.17){\makebox(0.00,0.00){ $ {\tilde{\alpha}}_{3}$ }}
\put(737.92,184.17){\makebox(0.00,0.00){$ \phi {\tilde{\alpha}}_{1}$}}
\put(708.17,149.59){\makebox(0.00,0.00){ $ \phi {\tilde{\alpha}}_{2}$}}
\put(647.50,165.00){\makebox(0.00,0.00){ $ \phi {\tilde{\alpha}}_{3}$ }}
\put(596.00,165.00){\makebox(0.00,0.00){${\tilde{\alpha}}_2$}}
\put(555.00,165.00){\makebox(0.00,0.00){${ \tilde{ \alpha}}_1$}}
\put(689.00,146.33){\line(1,-1){22.67}}
\put(711.67,176.67){\line(-1,-1){23.00}}
\put(650.00,150.00){\line(1,0){30.00}}
\put(630.00,150.00){\line(1,0){10.00}}
\put(600.00,150.00){\line(1,0){30.00}}
\put(560.00,150.00){\line(1,0){30.00}}
\put(715.00,120.00){\circle*{10.00}}
\put(715.00,180.00){\circle*{10.00}}
\put(685.00,150.00){\circle*{10.00}}
\put(645.00,150.00){\circle*{10.00}}
\put(555.00,150.00){\circle*{10.00}}
\put(595.00,150.00){\circle*{10.00}}
\end{picture}

\noindent The Cartan matrix (\ref{KN}) of $D_{6}$ then reads 
\begin{equation}
K=\left( 
\begin{array}{rrrrrr}
2 & -1 & 0 & 0 & 0 & 0 \\ 
-1 & 2 & 0 & 0 & 0 & -1 \\ 
0 & 0 & 2 & 0 & -1 & 0 \\ 
0 & 0 & 0 & 2 & -1 & 0 \\ 
0 & 0 & -1 & -1 & 2 & -1 \\ 
0 & -1 & 0 & 0 & -1 & 2%
\end{array}
\right) =R\left( 
\begin{array}{rr}
\tilde{K} & \phi \tilde{K} \\ 
\phi \tilde{K} & \phi ^{2}\tilde{K}%
\end{array}
\right) ,  \label{kk}
\end{equation}
where we also exhibit relation (\ref{KN}). Noting further that the Cartan
matrix of $H_{3}$ is 
\begin{equation}
\tilde{K}=\left( 
\begin{array}{rrr}
2 & -1 & 0 \\ 
-1 & 2 & -\phi \\ 
0 & -\phi & 2%
\end{array}
\right) ,  \label{kt}
\end{equation}
the relation (\ref{KK}) is read off directly by comparing (\ref{kk}) with (%
\ref{kt}). The Coxeter numbers are $h=\tilde{h}=10$ and the set of exponents
separate according to (\ref{SSS}) into 
\begin{equation}
\{1,3,5,5,7,9\}=\{1,5,9\}\cup \{3,5,7\}.
\end{equation}
Let us now see how $\omega $ acts on the orbits $\Omega _{i}$ and how the
map (\ref{PO}) is realized. We choose $\sigma =$ $\sigma _{1}\sigma
_{4}\sigma _{3}\sigma _{6}\sigma _{2}\sigma _{5}$ and $\tilde{\sigma}=$ $%
\tilde{\sigma}_{1}\tilde{\sigma}_{3}\tilde{\sigma}_{2}$ for the Coxeter
element of $D_{6}$ and $H_{3}$, respectively. The corresponding orbits $%
\Omega _{i}$ and $\tilde{\Omega}_{i}$ are computed by successive actions of $%
\sigma $ and $\tilde{\sigma}$, respectively, on the simple roots. One
realizes that the map $\omega $ relates them indeed as specified in (\ref{PO}%
). See appendix A for the explicit computation of the orbits.

Once more we may check how the root system $\Delta $ can be mapped into $%
\tilde{\Delta}\cup \phi \tilde{\Delta}$ orbit by orbit. First we relate $%
\mathcal{W}$ to $\mathcal{\tilde{W}}$ and verify (\ref{ss}) for the
generating relations of $\mathcal{\tilde{W}}$, which for $H_{3}$ are known
to be $\tilde{\sigma}_{i}^{2}=\tilde{\sigma}_{i+\tilde{\ell}}^{2}=1$ for $%
i=1,2,3$ together with 
\begin{equation}
(\tilde{\sigma}_{2}\tilde{\sigma}_{3})^{5}=(\tilde{\sigma}_{1}\tilde{\sigma}%
_{2})^{3}=(\tilde{\sigma}_{1}\tilde{\sigma}_{3})^{2}=1~.  \label{genH3}
\end{equation}
We verify that (\ref{genH3}) corresponds to 
\begin{equation}
(\sigma _{2}\sigma _{5}\sigma _{3}\sigma _{6})^{5}=(\sigma _{1}\sigma
_{4}\sigma _{2}\sigma _{5})^{3}=(\sigma _{1}\sigma _{4}\sigma _{3}\sigma
_{6})^{2}=1~.
\end{equation}

\noindent By the same reasoning as in the $H_{2}$-case it also follows that $%
(\sigma _{i}\sigma _{i+\tilde{\ell}})^{2}=1$.

\subsection{The embedding $H_{4}\hookrightarrow E_{8}$}

Also in this case we first fix our conventions for naming the simple roots
in order to guarantee that we can realize the map $\omega $ as defined in (%
\ref{pi2}). Here we label according to the Coxeter graph

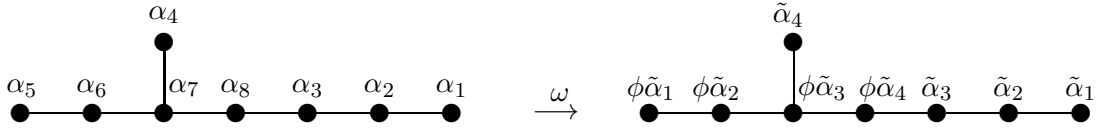
\unitlength=0.680000pt 
\begin{picture}(370.00,107.58)(220.00,0.00)
\put(445.00,52.00){\makebox(0.00,0.00){$\alpha_1$}}
\put(405.00,52.00){\makebox(0.00,0.00){$\alpha_2$}}
\put(365.00,52.00){\makebox(0.00,0.00){$\alpha_3$}}
\put(325.00,52.00){\makebox(0.00,0.00){$\alpha_8$}}
\put(296.25,52.17){\makebox(0.00,0.00){${\alpha}_7$}}
\put(285.00,93.00){\makebox(0.00,0.00){${\alpha}_4$}}
\put(245.00,52.00){\makebox(0.00,0.00){${\alpha}_6$}}
\put(206.00,52.00){\makebox(0.00,0.00){${\alpha}_5$}}
\put(410.00,38.00){\line(1,0){30.00}}
\put(370.00,38.00){\line(1,0){30.00}}
\put(330.00,38.00){\line(1,0){30.00}}
\put(285.33,72.67){\line(0,-1){31.00}}
\put(290.00,38.00){\line(1,0){30.00}}
\put(250.00,38.00){\line(1,0){30.00}}
\put(210.00,38.00){\line(1,0){30.00}}
\put(285.00,78.00){\circle*{10.00}}
\put(245.00,38.00){\circle*{10.00}}
\put(325.00,38.00){\circle*{10.00}}
\put(365.00,38.00){\circle*{10.00}}
\put(405.00,38.00){\circle*{10.00}}
\put(445.00,38.00){\circle*{10.00}}
\put(285.00,38.00){\circle*{10.00}}
\put(205.00,38.00){\circle*{10.00}}
\put(500.00,45.00){$\omega$}
\put(490.00,35.00){$ \longrightarrow $}
\put(795.00,52.00){\makebox(0.00,0.00){ $\tilde{\alpha}_1$ }}
\put(755.00,52.00){\makebox(0.00,0.00){ $\tilde{\alpha}_2$ }}
\put(715.00,52.00){\makebox(0.00,0.00){ $\tilde{\alpha}_3$ }}
\put(685.00,52.00){\makebox(0.00,0.00){ $ \phi \tilde{\alpha}_4$ }}
\put(651.25,52.17){\makebox(0.00,0.00){ $\phi \tilde{\alpha}_3$ }}
\put(635.00,93.00){\makebox(0.00,0.00){$\tilde{\alpha}_4$ }}
\put(595.00,52.00){\makebox(0.00,0.00){$ \phi \tilde{\alpha}_2$ }}
\put(556.00,52.00){\makebox(0.00,0.00){ $\phi \tilde{\alpha}_1$ }}
\put(760.00,38.00){\line(1,0){30.00}}
\put(720.00,38.00){\line(1,0){30.00}}
\put(680.00,38.00){\line(1,0){30.00}}
\put(635.33,72.67){\line(0,-1){31.00}}
\put(640.00,38.00){\line(1,0){30.00}}
\put(600.00,38.00){\line(1,0){30.00}}
\put(560.00,38.00){\line(1,0){30.00}}
\put(635.00,78.00){\circle*{10.00}}
\put(595.00,38.00){\circle*{10.00}}
\put(675.00,38.00){\circle*{10.00}}
\put(715.00,38.00){\circle*{10.00}}
\put(755.00,38.00){\circle*{10.00}}
\put(795.00,38.00){\circle*{10.00}}
\put(635.00,38.00){\circle*{10.00}}
\put(555.00,38.00){\circle*{10.00}}
\end{picture}

\noindent The corresponding Cartan matrix of $E_{8}$ together with its
construction from inner products in $\tilde{\Delta}$ in agreement with (\ref%
{KN}) is 
\begin{equation}
K=\left( 
\begin{array}{rrrrrrrr}
2 & -1 & 0 & 0 & 0 & 0 & 0 & 0 \\ 
-1 & 2 & -1 & 0 & 0 & 0 & 0 & 0 \\ 
0 & -1 & 2 & 0 & 0 & 0 & 0 & -1 \\ 
0 & 0 & 0 & 2 & 0 & 0 & -1 & 0 \\ 
0 & 0 & 0 & 0 & 2 & -1 & 0 & 0 \\ 
0 & 0 & 0 & 0 & -1 & 2 & -1 & 0 \\ 
0 & 0 & 0 & -1 & 0 & -1 & 2 & -1 \\ 
0 & 0 & -1 & 0 & 0 & 0 & -1 & 2%
\end{array}
\right) =R\left( 
\begin{array}{rr}
\tilde{K} & \phi \tilde{K} \\ 
\phi \tilde{K} & \phi ^{2}\tilde{K}%
\end{array}
\right) .  \label{34}
\end{equation}
The Cartan matrix of $H_{4}$ reads 
\begin{equation}
\tilde{K}=\left( 
\begin{array}{rrrr}
2 & -1 & 0 & 0 \\ 
-1 & 2 & -1 & 0 \\ 
0 & -1 & 2 & -\phi \\ 
0 & 0 & -\phi & 2%
\end{array}
\right) .  \label{35}
\end{equation}
Now the Coxeter numbers are $h=\tilde{h}=30$ and the set of exponents
separate according to (\ref{SSS}) into 
\begin{equation}
\{1,7,11,13,17,19,23,29\}=\{1,11,19,29\}\cup \{7,13,17,23\}.
\end{equation}
In order to see how $\omega $ acts on the orbits $\Omega _{i}$ and how the
map (\ref{PO}) is realized, we choose $\sigma =$ $\sigma _{1}\sigma
_{5}\sigma _{3}\sigma _{7}\sigma _{2}\sigma _{6}\sigma _{4}\sigma _{8}$ and $%
\tilde{\sigma}=$ $\tilde{\sigma}_{1}\tilde{\sigma}_{3}\tilde{\sigma}_{2}%
\tilde{\sigma}_{4}$ for the Coxeter element of $E_{8}$ and $H_{4}$,
respectively. We may then compute the corresponding orbits $\Omega _{i}$, $%
\tilde{\Omega}_{i}$ by successive actions of $\sigma $, $\tilde{\sigma}$ and
the simple roots and realize that the map $\omega $ relates them indeed as
specified in (\ref{PO}). See appendix for the orbits. The individual
reflections are related as (\ref{ss}) and the generating relations for the
Coxeter group are 
\begin{equation}
\left( \tilde{\sigma}_{6}\tilde{\sigma}_{7}\right) ^{5}=(\tilde{\sigma}_{4}%
\tilde{\sigma}_{5})^{3}=(\tilde{\sigma}_{4}\tilde{\sigma}_{6})^{3}=1~.
\end{equation}
which correspond to 
\begin{equation}
(\sigma _{2}\sigma _{6}\sigma _{3}\sigma _{7})^{5}=(\sigma _{4}\sigma
_{8}\sigma _{1}\sigma _{5})^{3}=(\sigma _{4}\sigma _{8}\sigma _{2}\sigma
_{6})^{3}=1~.
\end{equation}

\noindent By the same reasoning as in the $H_{2}$-case it also follows that $%
(\sigma _{i}\sigma _{i+\tilde{\ell}})^{2}=1$ .

\section{Affine Toda field theories}

\subsection{Generalities}

In this section we will demonstrate how one may construct an affine Toda
field theory related to a non-crystallographic Coxeter group $\mathcal{%
\tilde{W}}$ from a theory related to a crystallographic Coxeter group $%
\mathcal{W}$ by means of the discussed embedding. We start by taking $G$ to
be a Lie group with $H\subset G$ a maximal Torus and $\mathbf{h}$ its Cartan
subalgebra. Then the affine Toda field theory Lagrangian can be expressed as 
\cite{Mass1,Mass2} 
\begin{equation}
\mathcal{L}=\frac{1}{\beta ^{2}}\limfunc{Tr}\left( \frac{1}{2}\ \partial
_{\mu }g^{-1}\partial ^{\mu }g-m^{2}gEg^{-1}E^{\dag }\right) ,\;  \label{L}
\end{equation}
where $g=\exp (\beta \Phi )\in H$, $\beta $ is a coupling constant and $m~$a
mass scale. The regular element $E=\eta \cdot h^{\prime }$ with conjugate $%
E^{\dag }$ can be expanded in the Cartan subalgebra in apposition $%
h_{i}^{\prime }\in H^{\prime }$ \cite{Kost}. We normalize the trace
according to the Cartan-Weyl basis, that is $\limfunc{Tr}\left( E_{\alpha
}E_{-\alpha }\right) =1$. Notice that one can not write down a Lagrangian of
the type (\ref{L}) and relate it directly to a non-crystallographic Coxeter
group, since there is no proper Lie group and Lie algebra associated to them.

Conventionally one introduces $\ell $ scalar fields $\phi _{i}$ by expanding
the field $\Phi $ in the Cartan subalgebra $\Phi =\sum_{i=1}^{\ell }\phi
_{i}H_{i}$ with Cartan Weyl generators $H_{i}\in \mathbf{h}$ \cite%
{Mass1,Mass2}. Developing now the Lagrangian (\ref{L}) in powers of the
coupling constant, it follows that the term of zeroth order in $\beta $,
i.e. the quadratic term in $\phi _{i}$ becomes 
\begin{equation}
-\frac{1}{2}\sum_{i=1}^{\ell }m_{i}^{2}~|\phi _{i}|^{2}=-\frac{1}{2}%
m^{2}\sum_{i=1}^{\ell }|\eta \cdot \alpha _{i}|^{2}~|\phi _{i}|^{2},
\label{2}
\end{equation}
such that the mass of particle $i$ can be identified as $m_{i}=m|\eta \cdot
\alpha _{i}|$. Here $\eta $ is the eigenvector of the Coxeter element with
eigenvalue $\exp (2\pi i/h)$. Proceeding to the first order term in $\beta $%
, the constant in front of the cubic terms in the fields divided by the
symmetry factor $3!$ is taken to define the three-point coupling. It is
computed \cite{Mass2} to 
\begin{equation}
C_{ijk}=\frac{4\beta m^{2}}{\sqrt{-h}}\varepsilon _{ijq}\Delta _{ijk}
\label{C}
\end{equation}
where $\Delta _{ijk}=\sqrt{s(s-m_{i})(s-m_{j})(s-m_{k})}$ with $%
s=(m_{i}+m_{j}+m_{k})/2$ \ is Heron's formula for the area of the triangle
formed by the masses of the three fusing particles $i,j,k$. The structure
constants $\varepsilon _{ijq}$ result from the Lie algebraic commutator $%
[E_{\gamma _{i}},E_{\sigma ^{q}\gamma _{j}}]=\varepsilon _{ijq}E_{\gamma
_{i}+\sigma ^{q}\gamma _{j}}=\varepsilon _{ijq}E_{\sigma ^{\tilde{q}}\gamma
_{\bar{k}}}$ and for simply laced Lie algebras are normalized to $%
\varepsilon _{ijq}=\pm 1,0$. The $\gamma _{i}=c_{i}\alpha _{i}$ are the
coloured simple roots introduced in section 2.2. In other words the
three-point coupling $C_{ijk}$ is non-vanishing if and only if the
commutator of step operators related to roots in the orbits $\Omega _{i}$
and $\Omega _{j}$ is non-zero. Through this reasoning the fusing rule \cite%
{PD1} 
\begin{equation}
C_{ijk}\neq 0\qquad \Leftrightarrow \qquad \gamma _{i}+\sigma ^{q}\gamma
_{j}+\sigma ^{p}\gamma _{k}=0.  \label{Frule}
\end{equation}
can be related to the ATFT-Lagrangian.

Having computed these classical quantities one may ask the question towards
a formulation of the corresponding quantum field theory. A first glimpse of
its nature is obtained from a perturbative computation of the mass
corrections. For ATFT the logarithmic divergencies of the self-energy
corrections can be removed simply by normal ordering \cite{CM1,DDV}. Then
the one loop corrections to the mass of the particle $k$ are just the finite
contributions resulting from a bubble graph, which were found \cite%
{BCDS2,BCDS,DDV} to be 
\begin{equation}
\delta m_{k}^{2}=-i\frac{\sum\nolimits_{ij}(m_{k}^{2})}{(2\pi )^{2}},
\label{dm}
\end{equation}
where the sum extends over all intermediate contributions 
\begin{equation}
\sum\nolimits_{ij}(m_{k}^{2})=i\pi \frac{C_{ijk}^{2}(\pi -\theta _{ij}^{k})}{%
m_{i}m_{j}\sin \theta _{ij}^{k}},  \label{ren}
\end{equation}
with $\theta _{ij}^{k}$ being the fusing angle for the process $%
i+j\rightarrow \bar{k}$.

One should note that instead of the Lie algebraic formulation for the affine
Toda field theory Lagrangian (\ref{L}), one can in principle also start with
a Lagrangian for which the trace in (\ref{L}) is already computed in the
adjoint representation, such that it involves only roots rather than Lie
algebraic quantities 
\begin{equation}
\mathcal{L}=\frac{1}{2}\ \partial _{\mu }\Phi \partial ^{\mu }\Phi -\frac{%
m^{2}}{\beta ^{2}}\sum\limits_{i=0}^{\ell }n_{i}\exp (\beta \alpha _{i}\cdot
\Phi ).  \label{L2}
\end{equation}
Here $\alpha _{0}=$ $-\sum\nolimits_{i=0}^{\ell }n_{i}\alpha _{i}$ is the
negative of the highest root and $n_{i}$ the Kac labels. In this case the
expansion of the Lagrangian in $\beta $ yields 
\begin{equation}
\left( M^{2}\right) _{ij}=m^{2}\sum\limits_{p=0}^{\ell }n_{p}\alpha
_{p}^{i}\alpha _{p}^{j}\qquad \text{and\qquad }C_{ijk}=\beta
m^{2}\sum\limits_{p=0}^{\ell }n_{p}\alpha _{p}^{i}\alpha _{p}^{j}\alpha
_{p}^{k}.  \label{MC}
\end{equation}
In most cases the two formulations are equivalent, however one is often more
advantageous than the other. For instance, (\ref{L2}) does not allow for a
generic case independent treatment as it relies on a special choice of the
basis for the roots $\alpha _{i}$ needed to ensure that the mass matrix
becomes diagonal. In addition (\ref{L}) yields an explanation \cite%
{Mass1,Mass2} for the fact that the masses of ATFT organize into the
Perron-Frobenius vector of the Cartan matrix and a generic derivation \cite%
{Mass2} of the fusing rule (\ref{Frule}). For our present purposes \ it is
important to note that (\ref{L}) relies on the existence of a Lie algebra,
whereas (\ref{L2}) only requires a root system. Thus in principle we could
write down a Lagrangian of the type (\ref{L2}) for non-crystallographic
Coxeter groups, but the formulation of an equivalent Lie algebraic version
would be impossible. In addition, classical integrability for ATFT is
established by means of a Lax pair formulation in terms of Lie algebraic
quantities \cite{OP1,Integ}. Therefore it would remain an open issue whether
for non-crystallographic Coxeter groups (\ref{L2}) corresponds to integrable
models or not. Here our formulation will not be a Lagrangian of the form (%
\ref{L2}) involving a non-crystallographic root system, but rather a reduced
system which exploits the previously discussed embedding structure with
regard to the Cartan subalgebra in apposition. The crystallographic Lie
algebra structure is preserved in the reduction procedure, from which we can
see immediately that the new theories also possess Lax pairs ensuring the
classical integrability.

\subsection{Reduction from crystallographic to non-crystallographic ATFT}

Somewhat analogue to the folding procedure \cite{DIO}, we may now alter the
above expansion of $\Phi $ in order to define a new theory, which in this
case only contains $\tilde{\ell}=\ell /2$ scalar fields $\tilde{\varphi}_{i}$%
. This is achieved by expanding 
\begin{equation}
\Phi =\sum_{i=1}^{\ell }\mu (\varphi _{i})H_{i},~\quad ~~\text{with \ }\mu
(\varphi _{i})=\frac{1}{\phi \sqrt{3}}\left\{ 
\begin{array}{ll}
\tilde{\varphi}_{i} & \text{for\quad }1\leq i\leq \tilde{\ell}=\ell /2 \\ 
\phi \tilde{\varphi}_{i-\tilde{\ell}}\qquad & \text{for\quad }\tilde{\ell}%
<i\leq \ell .%
\end{array}
\right. ~  \label{shift}
\end{equation}
Here the map $\mu $ is inspired by the previously introduced map $\tilde{%
\omega}$ (\ref{at}). At this point one could have also defined $\mu $ in
such a way that it multiplies the fields $\varphi _{i}$ for $\tilde{\ell}%
<i\leq \ell $ by $\phi ^{-1}$. However, the choice (\ref{shift}) is distinct
by the structure emerging below for the three-point coupling. Other
possibilities are also conceivable. Having an alternative expansion for the
fields $\Phi $ as in (\ref{shift}), we re-consider first the quadratic term (%
\ref{2}) in the expansion of the Lagrangian, which now becomes 
\begin{equation}
-\frac{1}{2}\sum_{i=1}^{\ell }m_{i}^{2}~|\mu (\tilde{\varphi}_{i})|^{2}=-%
\frac{1}{2}\sum_{i=1}^{\tilde{\ell}}\frac{1}{3}(\phi ^{-2}m_{i}^{2}+m_{i+%
\tilde{\ell}}^{2})~|(\tilde{\varphi}_{i})|^{2}.
\end{equation}
We take this identity as the defining relation for the masses of the $\tilde{%
\ell}$ new scalar fields $\tilde{\phi}_{i}$ 
\begin{equation}
\tilde{m}_{i}^{2}=\frac{1}{3}(\phi ^{-2}m_{i}^{2}+m_{i+\tilde{\ell}%
}^{2})=m_{i}^{2}~.  \label{m}
\end{equation}
In (\ref{m}) we made use of the fact that the masses $m_{i}$ can be
identified as the components of the Perron-Frobenius eigenvector of $K$,
that is $m_{i}=(y_{1})_{i}$. Subsequently we employ (\ref{yn}) which implies
that $m_{i}=\phi m_{i+\tilde{\ell}}$ and then $\phi ^{-2}+\phi ^{2}=3$. We
also notice the fact that the masses $\tilde{m}_{i}$ admit a
\textquotedblleft genuine\textquotedblright\ interpretation as belonging to
an affine Toda field theory related to a non-crystallographic Coxeter group,
since (\ref{yn}) also implies that the masses $\tilde{m}_{i}$ are the
components of the Perron-Frobenius eigenvector of $\tilde{K}(\phi )$.

\noindent Similarly we proceed further and read off the next order term in $%
\beta $ to define a new three-point coupling $\tilde{C}_{ijk}$. We compute 
\begin{eqnarray}
\tilde{C}_{ijk} &=&\phi ^{3}C_{(i+\ell )(j+\tilde{\ell})(k+\tilde{\ell}%
)}+\phi ^{2}\left( C_{i(j+\tilde{\ell})(k+\tilde{\ell})}+C_{(i+\tilde{\ell}%
)j(k+\tilde{\ell})}+C_{(i+\tilde{\ell})(j+\tilde{\ell})k}\right)  \notag \\
&&+\phi \left( C_{(i+\tilde{\ell})jk}+C_{i(j+\tilde{\ell})k}+C_{ij(k+\tilde{%
\ell})}\right) +C_{ijk},  \label{CT}
\end{eqnarray}
for $1\leq i,j,k,\leq \tilde{\ell}$. The identification of the Coxeter
element (\ref{ss}) translates the fusing rule (\ref{Frule}) into 
\begin{eqnarray}
\tilde{C}_{ijk}\neq 0\qquad &\Leftrightarrow &\qquad \omega \gamma _{i}+%
\tilde{\sigma}^{q}\omega \gamma _{j}+\tilde{\sigma}^{p}\omega \gamma _{k}=0,
\label{FU} \\
&\Leftrightarrow &\qquad \phi ^{t_{i}}\tilde{\gamma}_{i}+\phi ^{t_{j}}\tilde{%
\sigma}^{q}\tilde{\gamma}_{j}+\phi ^{t_{k}}\tilde{\sigma}^{p}\tilde{\gamma}%
_{k}=0,  \label{FU2}
\end{eqnarray}
where the $t_{i}$ are the integers introduced in equation (\ref{KN}). The
relation (\ref{FU2}) is entirely expressed in quantities belonging to the
non-crystallographic Coxeter group. In comparison with the fusing rule
related to ATFT for simply laced Lie algebras (\ref{Frule}), it had to be
\textquotedblleft $\phi $-deformed\textquotedblright , somewhat similar to
the q-deformed versions of the fusing rule needed for ATFT associated with
non-simply laced Lie algebras \cite{q2}. Note that besides the two
non-equivalent solutions related as $q\rightarrow h+1/2(c_{i}-c_{j})-q$ and $%
p\rightarrow h+1/2(c_{i}-c_{j})-p$ \cite{FO,q2}, in (\ref{FU2}) there could
be more solutions corresponding to different values of the integers $%
t_{i},t_{j},t_{k}$.

Alternatively, we may compute the masses and three-point couplings from the
Lagrangian (\ref{L2}). Using the expansion (\ref{shift}) in (\ref{L2})
yields the same quantities. We obtain now 
\begin{equation}
\left( \tilde{M}^{2}\right) _{ij}=m^{2}\sum\limits_{p=0}^{\ell }n_{p}\hat{%
\alpha}_{p}^{i}\hat{\alpha}_{p}^{j}\qquad \text{and\qquad }\tilde{C}%
_{ijk}=\beta m^{2}\sum\limits_{p=0}^{\ell }n_{p}\hat{\alpha}_{p}^{i}\hat{%
\alpha}_{p}^{j}\hat{\alpha}_{p}^{k}.
\end{equation}
where we have folded the $\ell $-components of the root $\alpha _{p}^{i}$
with $1\leq i\leq \ell $ into an $\tilde{\ell}$-component vector 
\begin{equation}
\hat{\alpha}_{p}^{i}=\frac{1}{3}(\phi ^{-1}\alpha _{p}^{i}+\alpha _{p}^{i+%
\tilde{\ell}})~,
\end{equation}
for $1\leq i\leq \tilde{\ell}$. We stress that this is not the same as
writing down (\ref{L2}) for a non-crystallographic root system, since it
still involves $\ell $ roots, albeit now represented in $\mathbb{R}^{\tilde{%
\ell}}$. Having computed the three-point couplings for the reduced theory,
we may compute the mass renormalisation from (\ref{ren}) as 
\begin{equation}
\sum\nolimits_{ij}(\tilde{m}_{k}^{2})=i\pi \frac{\tilde{C}_{ijk}^{2}(\pi
-\theta _{ij}^{k})}{\tilde{m}_{i}\tilde{m}_{j}\sin \theta _{ij}^{k}}.
\label{mt}
\end{equation}
This will shed light on the possible form of the scattering matrix for ATFT
related to non-crystallographic Coxeter groups.

\subsection{From $A_{4}$ to $H_{2}$-affine Toda field theory}

Let us now make the above general formulae more explicit. When ignoring the
overall mass scale, the masses of $A_{4}$-ATFT can be brought into the
simple form 
\begin{equation}
m_{1}=m_{2}=1\qquad \text{and\qquad }m_{3}=m_{4}=\phi .
\end{equation}
Keeping the same normalization for the overall mass scale, the identity (\ref%
{m}) yields for the classical masses of the $H_{2}$-ATFT 
\begin{equation}
\tilde{m}_{1}=m_{1}\qquad \text{and\qquad }\tilde{m}_{2}=m_{2}.
\end{equation}
According to (\ref{C}) and (\ref{CT}) we then compute the three point
couplings $C_{ijk}$ and $\tilde{C}_{ijk}$ together with their corresponding
fusing rules, which result from the tables provided in section 2.3 
\begin{equation}
\begin{array}{|l|l|l|}
\hline
C_{113}=\frac{4i}{\sqrt{5}}\Delta _{113} & \gamma _{1}+\sigma \gamma
_{1}+\sigma ^{3}\gamma _{3}=0 & \tilde{C}_{111}=3\phi C_{113}=\frac{4i}{%
\sqrt{5}}\sqrt{9+12\phi }\tilde{\Delta}_{111} \\ \hline
C_{144}=\phi C_{113} & \gamma _{1}+\sigma ^{2}\gamma _{4}+\sigma ^{4}\gamma
_{4}=0 & \tilde{C}_{122}=\phi ^{3}C_{113} \\ \hline
C_{224}=-C_{113} & \gamma _{2}+\sigma \gamma _{2}+\sigma ^{3}\gamma _{4}=0 & 
\tilde{C}_{222}=-\tilde{C}_{111} \\ \hline
C_{233}=-\phi C_{113} & \gamma _{2}+\sigma \gamma _{3}+\sigma ^{3}\gamma
_{3}=0 & \tilde{C}_{211}=-\tilde{C}_{122}. \\ \hline
\end{array}%
\end{equation}
We did not report the factor $\beta m^{2}$ in each of the couplings. Here $%
\tilde{\Delta}_{ijk}$ is the area of the triangle bounded by the masses $%
\tilde{m}_{i},\tilde{m}_{j},\tilde{m}_{k}$. Note that now the factor of
proportionality between $|\tilde{C}_{ijk}|$ and $\tilde{\Delta}_{ijk}$ is no
longer universal as in (\ref{C}). This is sufficient information to carry
out the perturbation theory up to order $\beta ^{2}$. The mass corrections
to $\tilde{m}_{1}$ according to (\ref{mt}) are computed by summing the
convergent contributions to the one-loop corrections, which up to the
symmetry factors are

\unitlength=0.680000pt 
\begin{picture}(237.92,80.00)(190.00,115.00)

\put(200.0,150.00){\makebox(0.00,0.00){$\Sigma (\tilde m_1^2) \quad = $}}

\put(250.00,150.00){\line(1,0){40.00}}
\put(350.00,150.00){\line(1,0){40.00}}
\qbezier(290.00,150.00)(320.00,175.00)(350.00,150.00)
\qbezier(290.00,150.00)(320.00,125.00)(350.00,150.00)
\put(270.0,140.00){\makebox(0.00,0.00){$\tilde 1$}}
\put(370.0,140.00){\makebox(0.00,0.00){$\tilde 1$}}
\put(320.0,175.00){\makebox(0.00,0.00){$\tilde 1$}}
\put(320.0,125.00){\makebox(0.00,0.00){$\tilde 1$}}
\put(420.0,150.00){\makebox(0.00,0.00){$+$}}
\put(450.00,150.00){\line(1,0){40.00}}
\put(550.00,150.00){\line(1,0){40.00}}
\qbezier(490.00,150.00)(520.00,175.00)(550.00,150.00)
\qbezier(490.00,150.00)(520.00,125.00)(550.00,150.00)
\put(470.0,140.00){\makebox(0.00,0.00){$\tilde 1$}}
\put(570.0,140.00){\makebox(0.00,0.00){$\tilde 1$}}
\put(520.0,175.00){\makebox(0.00,0.00){$\tilde 1$}}
\put(520.0,125.00){\makebox(0.00,0.00){$\tilde 2$}}

\put(620.0,150.00){\makebox(0.00,0.00){$+$}}
\put(650.00,150.00){\line(1,0){40.00}}
\put(750.00,150.00){\line(1,0){40.00}}
\qbezier(690.00,150.00)(720.00,175.00)(750.00,150.00)
\qbezier(690.00,150.00)(720.00,125.00)(750.00,150.00)
\put(670.0,140.00){\makebox(0.00,0.00){$\tilde 1$}}
\put(770.0,140.00){\makebox(0.00,0.00){$\tilde 1$}}
\put(720.0,175.00){\makebox(0.00,0.00){$\tilde 2$}}
\put(720.0,125.00){\makebox(0.00,0.00){$\tilde 2$}}
\end{picture}

\noindent We omitted the usual arrows indicating the time direction and
assume throughout this paper that they all run to the right hand side. We
computed the symmetry factors by applying Wick's theorem and include them
into formula (\ref{mt}), such that 
\begin{equation}
\sum\nolimits_{ij}(\tilde{m}_{1}^{2})=i\pi \frac{\frac{\pi }{3}}{\sin \frac{%
2\pi }{3}}\left( 18\tilde{C}_{111}^{2}+36\tilde{C}_{112}^{2}+18\tilde{C}%
_{221}^{2}\right) ~.
\end{equation}
where we assume that the particles are conjugate to each other, that is $%
\bar{1}=2$, $\bar{2}=1$ where the bar indicates the anti-particle.
Similarly, the mass corrections to $\tilde{m}_{2}$ result from summing

\unitlength=0.680000pt 
\begin{picture}(237.92,80.00)(190.00,115.00)

\put(200.0,150.00){\makebox(0.00,0.00){$\Sigma (\tilde m_2^2) \quad = $}}

\put(250.00,150.00){\line(1,0){40.00}}
\put(350.00,150.00){\line(1,0){40.00}}
\qbezier(290.00,150.00)(320.00,175.00)(350.00,150.00)
\qbezier(290.00,150.00)(320.00,125.00)(350.00,150.00)
\put(270.0,140.00){\makebox(0.00,0.00){$\tilde 2$}}
\put(370.0,140.00){\makebox(0.00,0.00){$\tilde 2$}}
\put(320.0,175.00){\makebox(0.00,0.00){$\tilde 2$}}
\put(320.0,125.00){\makebox(0.00,0.00){$\tilde 2$}}
\put(420.0,150.00){\makebox(0.00,0.00){$+$}}
\put(450.00,150.00){\line(1,0){40.00}}
\put(550.00,150.00){\line(1,0){40.00}}
\qbezier(490.00,150.00)(520.00,175.00)(550.00,150.00)
\qbezier(490.00,150.00)(520.00,125.00)(550.00,150.00)
\put(470.0,140.00){\makebox(0.00,0.00){$\tilde 2$}}
\put(570.0,140.00){\makebox(0.00,0.00){$\tilde 2$}}
\put(520.0,175.00){\makebox(0.00,0.00){$\tilde 2$}}
\put(520.0,125.00){\makebox(0.00,0.00){$\tilde 1$}}

\put(620.0,150.00){\makebox(0.00,0.00){$+$}}
\put(650.00,150.00){\line(1,0){40.00}}
\put(750.00,150.00){\line(1,0){40.00}}
\qbezier(690.00,150.00)(720.00,175.00)(750.00,150.00)
\qbezier(690.00,150.00)(720.00,125.00)(750.00,150.00)
\put(670.0,140.00){\makebox(0.00,0.00){$\tilde 2$}}
\put(770.0,140.00){\makebox(0.00,0.00){$\tilde 2$}}
\put(720.0,175.00){\makebox(0.00,0.00){$\tilde 1$}}
\put(720.0,125.00){\makebox(0.00,0.00){$\tilde 1$}}
\end{picture}

\noindent which gives 
\begin{equation}
\sum\nolimits_{ij}(\tilde{m}_{2}^{2})=i\pi \frac{\frac{\pi }{3}}{\sin \frac{%
2\pi }{3}}\left( 18\tilde{C}_{222}^{2}+36\tilde{C}_{221}^{2}+18\tilde{C}%
_{112}^{2}\right) ~.
\end{equation}%
Assembling this according to (\ref{dm}) yields the important fact that the
classical mass ratios are conserved in the quantum field theory 
\begin{equation}
\frac{\delta m_{1}^{2}}{m_{1}^{2}}=\frac{\delta m_{2}^{2}}{m_{2}^{2}}=\frac{%
\sqrt{3}}{4}(5+3\phi )\tilde{C}_{111}^{2}~.
\end{equation}%
This means at first order perturbation theory the masses renormalise
equally. As the masses of both particles coincide and they undergo the same
fusing processes, they appear to be indistinguishable at this stage. This
possibly hints towards a non-diagonal scattering theory, that means a theory
in which backscattering is possible. However, it remains to be seen whether
there exist higher charges in this theory which make the particles distinct.

\subsection{From $D_{6}$ to $H_{3}$-affine Toda field theory}

The masses of $D_{6}$-ATFT are known for a long time and can be brought into
the form (\ref{mD}). Keeping the same normalization for the overall mass
scale, the identity (\ref{m}) yields for the classical masses of the $H_{3}$%
-ATFT 
\begin{equation}
\tilde{m}_{1}=\phi ^{-1},\qquad \tilde{m}_{2}=\sqrt{1+\phi ^{-2}},\qquad 
\tilde{m}_{3}=1.
\end{equation}%
According to (\ref{C}) and (\ref{CT}) we then compute the three point
couplings $C_{ijk}$ and $\tilde{C}_{ijk}$ together with their corresponding
fusing rules, which result from the tables provided in the appendix 
\begin{equation}
\begin{array}{|l|l|l|}
\hline
\begin{array}{l}
C_{112}=4i/\sqrt{10}\Delta _{112} \\ 
C_{442}=\phi ^{3}C_{112} \\ 
C_{445}=\phi ^{2}C_{112}%
\end{array}
& 
\begin{array}{l}
\gamma _{1}+\sigma \gamma _{1}+\sigma ^{6}\gamma _{2}=0 \\ 
\gamma _{4}+\sigma ^{3}\gamma _{4}+\sigma ^{7}\gamma _{2}=0 \\ 
\gamma _{4}+\sigma \gamma _{4}+\sigma ^{6}\gamma _{5}=0%
\end{array}
& \tilde{C}_{112}=(10\phi +7)\tilde{\Delta}_{112} \\ \hline
\begin{array}{l}
C_{665}=-\phi ^{5}C_{112} \\ 
C_{332}=\phi ^{3}C_{112} \\ 
C_{335}=\phi ^{2}C_{112}%
\end{array}
& 
\begin{array}{l}
\gamma _{6}+\sigma ^{3}\gamma _{6}+\sigma ^{7}\gamma _{5}=0 \\ 
\gamma _{3}+\sigma ^{3}\gamma _{3}+\sigma ^{7}\gamma _{2}=0 \\ 
\gamma _{3}+\sigma \gamma _{3}+\sigma ^{6}\gamma _{5}=0%
\end{array}
& \tilde{C}_{332}=-(1+5\phi )\tilde{\Delta}_{332} \\ \hline
\begin{array}{l}
C_{126}=\phi ^{2}C_{112} \\ 
C_{156}=\phi ^{3}C_{112}%
\end{array}
& 
\begin{array}{l}
\gamma _{1}+\sigma ^{2}\gamma _{2}+\sigma ^{6}\gamma _{6}=0 \\ 
\gamma _{1}+\sigma ^{4}\gamma _{5}+\sigma ^{8}\gamma _{6}=0%
\end{array}
& \tilde{C}_{123}=5.56758\tilde{\Delta}_{123} \\ \hline
\begin{array}{l}
C_{225}=(\phi ^{4}-1)C_{112} \\ 
C_{255}=(\phi -\phi ^{5})C_{112}%
\end{array}
& 
\begin{array}{l}
\gamma _{2}+\sigma ^{2}\gamma _{2}+\sigma ^{6}\gamma _{5}=0 \\ 
\gamma _{2}+\sigma ^{3}\gamma _{5}+\sigma ^{7}\gamma _{5}=0%
\end{array}
& \tilde{C}_{222}=-8.6253\tilde{\Delta}_{222} \\ \hline
C_{134}=\phi ^{2}C_{112} & \gamma _{1}+\sigma ^{3}\gamma _{3}+\sigma
^{7}\gamma _{4}=0 & \tilde{C}_{113}=2\phi ^{2}\tilde{\Delta}_{113} \\ \hline
C_{346}=\phi ^{3}C_{112} & \gamma _{3}+\sigma ^{8}\gamma _{4}+\sigma
^{4}\gamma _{6}=0 & \tilde{C}_{331}=2\phi ^{3}\tilde{\Delta}_{331}. \\ \hline
\end{array}%
\end{equation}%
Here we did not report the factor of $i\beta m^{2}4/\sqrt{10}$ in $\tilde{C}$%
. The fusing rules reduce according to (\ref{FU}), for instance 
\begin{eqnarray}
\gamma _{1}+\sigma \gamma _{1}+\sigma ^{6}\gamma _{2} &=&0~\Rightarrow ~%
\tilde{\gamma}_{1}+\tilde{\sigma}\tilde{\gamma}_{1}+\tilde{\sigma}^{6}\tilde{%
\gamma}_{2}=0,  \label{F1} \\
\gamma _{4}+\sigma ^{3}\gamma _{4}+\sigma ^{7}\gamma _{2} &=&0~\Rightarrow
~\phi \tilde{\gamma}_{1}+\phi \tilde{\sigma}^{3}\tilde{\gamma}_{1}+\tilde{%
\sigma}^{7}\tilde{\gamma}_{2}=0,  \label{F2} \\
\gamma _{4}+\sigma \gamma _{4}+\sigma ^{6}\gamma _{5} &=&0~\Rightarrow ~\phi 
\tilde{\gamma}_{1}+\phi \tilde{\sigma}\tilde{\gamma}_{1}+\phi \sigma
^{6}\gamma _{2}=0.  \label{F3}
\end{eqnarray}%
Note that we can construct the solution (\ref{F3}) trivially from (\ref{F1})
simply by multiplying it with $\phi $. However, (\ref{F2}) can not be
obtained from (\ref{F1}) or (\ref{F3}) in such a manner and has to be
regarded as independent.

As described in the previous section, we compute the mass renormalisation to 
\begin{eqnarray}
\sum\nolimits_{ij}(\tilde{m}_{1}^{2}) &=&i\pi \left( \frac{36\tilde{C}%
_{112}^{2}\frac{\pi }{10}}{\tilde{m}_{1}\tilde{m}_{2}\sin \frac{9\pi }{10}}+%
\frac{36\tilde{C}_{123}^{2}\hat{\phi}}{\tilde{m}_{2}\tilde{m}_{3}\sin (\pi -%
\hat{\phi})}+\frac{18\tilde{C}_{133}^{2}\frac{2\pi }{10}}{\tilde{m}_{3}%
\tilde{m}_{3}\sin \frac{8\pi }{10}}+\frac{36\tilde{C}_{113}^{2}\frac{2\pi }{%
10}}{\tilde{m}_{1}\tilde{m}_{3}\sin \frac{8\pi }{10}}\right) \\
~\sum\nolimits_{ij}(\tilde{m}_{2}^{2}) &=&i\pi \left( \frac{18\tilde{C}%
_{222}^{2}\frac{\pi }{3}}{\tilde{m}_{2}\tilde{m}_{2}\sin \frac{2\pi }{3}}+%
\frac{36\tilde{C}_{123}^{2}\frac{\pi }{2}}{\tilde{m}_{1}\tilde{m}_{3}\sin 
\frac{\pi }{2}}+\frac{18\tilde{C}_{112}^{2}\frac{8\pi }{10}}{\tilde{m}_{1}%
\tilde{m}_{1}\sin \frac{2\pi }{10}}+\frac{18\tilde{C}_{233}^{2}\frac{4\pi }{%
10}}{\tilde{m}_{3}\tilde{m}_{3}\sin \frac{6\pi }{10}}\right) \\
\sum\nolimits_{ij}(\tilde{m}_{3}^{2}) &=&i\pi \left( \frac{36\tilde{C}%
_{233}^{2}\frac{3\pi }{10}}{\tilde{m}_{2}\tilde{m}_{3}\sin \frac{7\pi }{10}}+%
\frac{36\tilde{C}_{123}^{2}\check{\phi}}{\tilde{m}_{2}\tilde{m}_{3}\sin (\pi
-\check{\phi})}+\frac{18\tilde{C}_{113}^{2}\frac{4\pi }{10}}{\tilde{m}_{1}%
\tilde{m}_{1}\sin \frac{6\pi }{10}}+\frac{36\tilde{C}_{133}^{2}\frac{4\pi }{%
10}}{\tilde{m}_{1}\tilde{m}_{3}\sin \frac{6\pi }{10}}\right) ~~~\quad
\end{eqnarray}
where we abbreviate $\hat{\phi}=\arctan \phi ^{-1}$, $\check{\phi}=\arctan
\phi $. From this it follows then that classical mass ratios are not
conserved in the quantum field theory 
\begin{equation}
\frac{\delta m_{1}^{2}}{m_{1}^{2}}=196.996\ldots \qquad \frac{\delta
m_{2}^{2}}{m_{2}^{2}}=647.392\ldots \qquad \text{and\qquad }\frac{\delta
m_{3}^{2}}{m_{3}^{2}}=924.343\ldots
\end{equation}
This means that the scattering matrix for $H_{3}$-ATFT can not be of the
simple form as for ATFT related to simply laced Lie algebras.

\subsection{From $E_{8}$ to $H_{4}$ affine Toda field theory}

The masses of $E_{8}$-ATFT in the form (\ref{m8}) indicate the underlying $%
H_{4}$ structure and (\ref{m}) yields for the classical masses for the $%
H_{4} $-ATFT 
\begin{equation}
\tilde{m}_{1}=m_{1},\qquad \tilde{m}_{2}=m_{2},\qquad \tilde{m}%
_{3}=m_{3},\qquad \tilde{m}_{4}=m_{4}.
\end{equation}
Similarly as in the previous section we compute the three point couplings $%
C_{ijk}$ and $\tilde{C}_{ijk}$ \ from (\ref{C}) and (\ref{CT}) together with
their corresponding fusing rules

\begin{tabular}{|l|l|l|}
\hline\hline
$%
\begin{array}{c}
C_{111}=\frac{4i}{\sqrt{30}}~0.433013 \\ 
C_{511}=\frac{4i}{\sqrt{30}}~0.475528 \\ 
C_{551}=\phi C_{511} \\ 
C_{555}=\phi ^{2}C_{111}%
\end{array}%
$ & $%
\begin{array}{c}
\gamma _{1}+\sigma ^{10}\gamma _{1}+\sigma ^{20}\gamma _{1}=0 \\ 
\gamma _{5}+\sigma ^{12}\gamma _{1}+\sigma ^{18}\gamma _{1}=0 \\ 
\gamma _{5}+\sigma ^{12}\gamma _{5}+\sigma ^{21}\gamma _{1}=0 \\ 
\gamma _{5}+\sigma ^{10}\gamma _{5}+\sigma ^{20}\gamma _{5}=0%
\end{array}%
$ & $\tilde{C}_{111}=31.3768~\tilde{\Delta}_{111}$ \\ \hline\hline
$%
\begin{array}{c}
C_{211}=\frac{4i}{\sqrt{30}}~0.103956 \\ 
C_{521}=-\phi ^{2}C_{321} \\ 
C_{655}=\phi ^{2}C_{211}%
\end{array}%
$ & $%
\begin{array}{c}
\gamma _{2}+\sigma ^{14}\gamma _{1}+\sigma ^{15}\gamma _{1}=0 \\ 
\gamma _{5}+\sigma ^{13}\gamma _{2}+\sigma ^{23}\gamma _{1}=0 \\ 
\gamma _{6}+\sigma ^{14}\gamma _{5}+\sigma ^{15}\gamma _{5}=0%
\end{array}%
$ & $\tilde{C}_{211}=37.1363~\tilde{\Delta}_{211}$ \\ \hline\hline
$%
\begin{array}{c}
C_{321}=-\frac{4i}{\sqrt{30}}~0.307324 \\ 
C_{631}=-C_{432}/\phi \\ 
C_{653}=C_{432} \\ 
C_{765}=-\phi ^{2}C_{321}%
\end{array}%
$ & $%
\begin{array}{c}
\gamma _{3}+\sigma ^{15}\gamma _{2}+\sigma ^{16}\gamma _{1}=0 \\ 
\gamma _{6}+\sigma ^{13}\gamma _{3}+\sigma ^{20}\gamma _{1}=0 \\ 
\gamma _{6}+\sigma ^{9}\gamma _{5}+\sigma ^{17}\gamma _{3}=0 \\ 
\gamma _{7}+\sigma ^{15}\gamma _{6}+\sigma ^{16}\gamma _{5}=0%
\end{array}%
$ & $\tilde{C}_{321}=22.6132~\tilde{\Delta}_{321}$ \\ \hline\hline
$%
\begin{array}{c}
C_{421}=\frac{4i}{\sqrt{30}}~0.972789 \\ 
C_{641}=-\phi ^{2}C_{321} \\ 
C_{861}=\phi ^{3}C_{321} \\ 
C_{865}=\phi ^{2}C_{421}%
\end{array}%
$ & $%
\begin{array}{c}
\gamma _{4}+\sigma ^{13}\gamma _{2}+\sigma ^{19}\gamma _{1}=0 \\ 
\gamma _{6}+\sigma ^{14}\gamma _{4}+\sigma ^{17}\gamma _{1}=0 \\ 
\gamma _{8}+\sigma ^{14}\gamma _{6}+\sigma ^{18}\gamma _{1}=0 \\ 
\gamma _{8}+\sigma ^{13}\gamma _{6}+\sigma ^{19}\gamma _{5}=0%
\end{array}%
$ & $\tilde{C}_{421}=9.92482~\tilde{\Delta}_{421}$ \\ \hline\hline
$%
\begin{array}{c}
C_{431}=-\frac{4i}{\sqrt{30}}~1.09848 \\ 
C_{831}=C_{421}/\phi \\ 
C_{871}=-C_{421} \\ 
C_{875}=-\phi ^{2}C_{431}%
\end{array}%
$ & $%
\begin{array}{c}
\gamma _{4}+\sigma ^{13}\gamma _{3}+\sigma ^{24}\gamma _{1}=0 \\ 
\gamma _{8}+\sigma ^{14}\gamma _{3}+\sigma ^{16}\gamma _{1}=0 \\ 
\gamma _{8}+\sigma ^{14}\gamma _{7}+\sigma ^{27}\gamma _{1}=0 \\ 
\gamma _{8}+\sigma ^{13}\gamma _{7}+\sigma ^{24}\gamma _{5}=0%
\end{array}%
$ & $\tilde{C}_{431}=8.65727~\tilde{\Delta}_{431}$ \\ \hline\hline
$%
\begin{array}{c}
C_{441}=\frac{4i}{\sqrt{30}}~1.17616 \\ 
C_{854}=-C_{421} \\ 
C_{885}=-\phi ^{2}C_{441}%
\end{array}%
$ & $%
\begin{array}{c}
\gamma _{4}+\sigma ^{13}\gamma _{4}+\sigma ^{21}\gamma _{1}=0 \\ 
\gamma _{8}+\sigma ^{13}\gamma _{5}+\sigma ^{16}\gamma _{4}=0 \\ 
\gamma _{8}+\sigma ^{13}\gamma _{8}+\sigma ^{21}\gamma _{5}=0%
\end{array}%
$ & $\tilde{C}_{441}=-14.4209~\tilde{\Delta}_{441}$ \\ \hline\hline
$%
\begin{array}{c}
C_{541}=C_{421}/\phi \\ 
C_{554}=-\phi ^{3}C_{321}%
\end{array}%
$ & $%
\begin{array}{c}
\gamma _{5}+\sigma ^{14}\gamma _{4}+\sigma ^{26}\gamma _{1}=0 \\ 
\gamma _{5}+\sigma ^{7}\gamma _{5}+\sigma ^{19}\gamma _{4}=0%
\end{array}%
$ & $%
\begin{array}{c}
\tilde{C}_{411}=5.35386%
\end{array}%
$ \\ \hline\hline
$%
\begin{array}{c}
C_{771}=-C_{432}%
\end{array}%
$ & $%
\begin{array}{c}
\gamma _{7}+\sigma ^{14}\gamma _{7}+\sigma ^{22}\gamma _{1}=0%
\end{array}%
$ & $%
\begin{array}{c}
\tilde{C}_{331}=-4.27444~\tilde{\Delta}_{331}%
\end{array}%
$ \\ \hline\hline
$%
\begin{array}{c}
C_{222}=-\frac{4i}{\sqrt{30}}~1.71313 \\ 
C_{622}=-\frac{4i}{\sqrt{30}}~1.88133 \\ 
C_{662}=\phi C_{622} \\ 
C_{666}=-\phi ^{2}C_{222}%
\end{array}%
$ & $%
\begin{array}{c}
\gamma _{2}+\sigma ^{10}\gamma _{2}+\sigma ^{20}\gamma _{2}=0 \\ 
\gamma _{6}+\sigma ^{18}\gamma _{2}+\sigma ^{12}\gamma _{2}=0 \\ 
\gamma _{6}+\sigma ^{12}\gamma _{6}+\sigma ^{21}\gamma _{2}=0 \\ 
\gamma _{6}+\sigma ^{10}\gamma _{6}+\sigma ^{20}\gamma _{6}=0%
\end{array}%
$ & $\tilde{C}_{222}=-9.1965~\tilde{\Delta}_{222}$ \\ \hline\hline
$%
\begin{array}{c}
C_{322}=-\frac{4i}{\sqrt{30}}~1.96731 \\ 
C_{762}=-C_{432} \\ 
C_{766}=-\phi ^{2}C_{322}%
\end{array}%
$ & $%
\begin{array}{c}
\gamma _{3}+\sigma ^{12}\gamma _{2}+\sigma ^{19}\gamma _{2}=0 \\ 
\gamma _{7}+\sigma ^{14}\gamma _{6}+\sigma ^{18}\gamma _{2}=0 \\ 
\gamma _{7}+\sigma ^{12}\gamma _{6}+\sigma ^{19}\gamma _{6}=0%
\end{array}%
$ & $\tilde{C}_{322}=3.75947~\tilde{\Delta}_{322}$ \\ \hline\hline
\end{tabular}

\begin{tabular}{|l|l|l|}
\hline\hline
$%
\begin{array}{c}
C_{522}=C_{432}/\phi \\ 
C_{652}=-\phi ^{3}C_{321}%
\end{array}%
$ & $%
\begin{array}{c}
\gamma _{5}+\sigma ^{10}\gamma _{2}+\sigma ^{21}\gamma _{2}=0 \\ 
\gamma _{6}+\sigma ^{12}\gamma _{5}+\sigma ^{17}\gamma _{2}=0%
\end{array}%
$ & $%
\begin{array}{c}
\tilde{C}_{122}=9.55253~\tilde{\Delta}_{122}%
\end{array}%
$ \\ \hline\hline
$%
\begin{array}{c}
C_{822}=\phi ^{2}C_{321}%
\end{array}%
$ & $%
\begin{array}{c}
\gamma _{8}+\sigma ^{14}\gamma _{2}+\sigma ^{16}\gamma _{2}=0%
\end{array}%
$ & $%
\begin{array}{c}
\tilde{C}_{422}=-0.683318~\tilde{\Delta}_{422}%
\end{array}%
$ \\ \hline\hline
$%
\begin{array}{c}
C_{432}=\frac{4i}{\sqrt{30}}~2.37859 \\ 
C_{832}=\phi ^{2}C_{431} \\ 
C_{863}=-\phi ^{3}C_{431} \\ 
C_{876}=\phi ^{2}C_{432}%
\end{array}%
$ & $%
\begin{array}{c}
\gamma _{4}+\sigma ^{11}\gamma _{3}+\sigma ^{22}\gamma _{2}=0 \\ 
\gamma _{8}+\sigma ^{12}\gamma _{3}+\sigma ^{19}\gamma _{2}=0 \\ 
\gamma _{8}+\sigma ^{11}\gamma _{6}+\sigma ^{19}\gamma _{3}=0 \\ 
\gamma _{8}+\sigma ^{11}\gamma _{7}+\sigma ^{22}\gamma _{6}=0%
\end{array}%
$ & $\tilde{C}_{432}=15.2555~\tilde{\Delta}_{432}$ \\ \hline\hline
$%
\begin{array}{c}
C_{732}=-C_{432}/\phi \\ 
C_{772}=-\phi ^{3}C_{431}%
\end{array}%
$ & $%
\begin{array}{c}
\gamma _{7}+\sigma ^{14}\gamma _{3}+\sigma ^{17}\gamma _{2}=0 \\ 
\gamma _{7}+\sigma ^{13}\gamma _{7}+\sigma ^{22}\gamma _{2}=0%
\end{array}%
$ & $%
\begin{array}{c}
\tilde{C}_{332}=2.68177~\tilde{\Delta}_{332}%
\end{array}%
$ \\ \hline\hline
$%
\begin{array}{c}
C_{333}=-\frac{4i}{\sqrt{30}}~3.78439 \\ 
C_{733}=\frac{4i}{\sqrt{30}}~4.15597 \\ 
C_{773}=\phi C_{733} \\ 
C_{777}=\phi ^{2}C_{333}%
\end{array}%
$ & $%
\begin{array}{c}
\gamma _{3}+\sigma ^{10}\gamma _{3}+\sigma ^{20}\gamma _{3}=0 \\ 
\gamma _{7}+\sigma ^{12}\gamma _{3}+\sigma ^{18}\gamma _{3}=0 \\ 
\gamma _{7}+\sigma ^{12}\gamma _{7}+\sigma ^{21}\gamma _{3}=0 \\ 
\gamma _{7}+\sigma ^{10}\gamma _{7}+\sigma ^{20}\gamma _{7}=0%
\end{array}%
$ & \multicolumn{1}{|c|}{$\tilde{C}_{333}=7.1965~\tilde{\Delta}_{333}$} \\ 
\hline\hline
$%
\begin{array}{c}
C_{433}=-\frac{4i}{\sqrt{30}}~3.24742 \\ 
C_{743}=-\phi ^{2}C_{431} \\ 
C_{877}=\phi ^{2}C_{433}%
\end{array}%
$ & $%
\begin{array}{c}
\gamma _{4}+\sigma ^{9}\gamma _{3}+\sigma ^{20}\gamma _{3}=0 \\ 
\gamma _{7}+\sigma ^{13}\gamma _{4}+\sigma ^{17}\gamma _{3}=0 \\ 
\gamma _{8}+\sigma ^{9}\gamma _{7}+\sigma ^{20}\gamma _{7}=0%
\end{array}%
$ & $\tilde{C}_{433}=-9.22437~\tilde{\Delta}_{433}$ \\ \hline\hline
$%
\begin{array}{c}
C_{553}=-C_{421}%
\end{array}%
$ & $%
\begin{array}{c}
\gamma _{5}+\sigma ^{4}\gamma _{5}+\sigma ^{17}\gamma _{3}=0%
\end{array}%
$ & $%
\begin{array}{c}
\tilde{C}_{113}=-2.5468%
\end{array}%
$ \\ \hline\hline
$%
\begin{array}{c}
C_{444}=-\frac{4i}{\sqrt{30}}~2.50428 \\ 
C_{844}=\frac{4i}{\sqrt{30}}~2.75016 \\ 
C_{884}=\phi C_{844} \\ 
C_{888}=-\phi ^{2}C_{444}%
\end{array}%
$ & $%
\begin{array}{c}
\gamma _{4}+\sigma ^{10}\gamma _{4}+\sigma ^{20}\gamma _{4}=0 \\ 
\gamma _{8}+\sigma ^{12}\gamma _{4}+\sigma ^{18}\gamma _{4}=0 \\ 
\gamma _{8}+\sigma ^{12}\gamma _{8}+\sigma ^{21}\gamma _{4}=0 \\ 
\gamma _{8}+\sigma ^{10}\gamma _{8}+\sigma ^{20}\gamma _{8}=0%
\end{array}%
$ & $\tilde{C}_{444}=29.3768~\tilde{\Delta}_{444}$ \\ \hline\hline
$%
\begin{array}{c}
C_{644}=\phi ^{2}C_{431}%
\end{array}%
$ & $%
\begin{array}{c}
\gamma _{6}+\sigma ^{11}\gamma _{4}+\sigma ^{19}\gamma _{4}=0%
\end{array}%
$ & $%
\begin{array}{c}
\tilde{C}_{244}=-2.13686~\tilde{\Delta}_{244}%
\end{array}%
$ \\ \hline\hline
$%
\begin{array}{c}
C_{744}=C_{421}/\phi \\ 
C_{874}=\phi ^{3}C_{431}%
\end{array}%
$ & $%
\begin{array}{c}
\gamma _{7}+\sigma ^{15}\gamma _{4}+\sigma ^{16}\gamma _{4}=0 \\ 
\gamma _{8}+\sigma ^{12}\gamma _{7}+\sigma ^{23}\gamma _{4}=0%
\end{array}%
$ & $%
\begin{array}{c}
\tilde{C}_{344}=-8.34233~\tilde{\Delta}_{344}%
\end{array}%
$ \\ \hline\hline
\end{tabular}

\medskip \noindent Also here we did not report the overall factor of $i\beta
m^{2}4/\sqrt{30}$ in $\tilde{C}$. Note that $\tilde{C}_{411}$ and $\tilde{C}%
_{311}$ have no classical mass triangle associated to them and therefore
yield no poles in the propagators. The nonvanishing one-loop contributions
are therefore \bigskip \medskip

\unitlength=0.680000pt 
\begin{picture}(237.92,115.00)(180.00,45.00)

\put(200.0,150.00){\makebox(0.00,0.00){$\Sigma (\tilde m_1^2) \quad = $}}

\put(250.00,150.00){\line(1,0){20.00}}
\put(310.00,150.00){\line(1,0){20.00}}
\qbezier(270.00,150.00)(290.00,170.00)(310.00,150.00)
\qbezier(270.00,150.00)(290.00,130.00)(310.00,150.00)
\put(260.0,140.00){\makebox(0.00,0.00){{\tiny $\tilde 1$}}}
\put(320.0,140.00){\makebox(0.00,0.00){\tiny $\tilde 1$}}
\put(290.0,170.00){\makebox(0.00,0.00){\tiny $\tilde 1$}}
\put(290.0,130.00){\makebox(0.00,0.00){\tiny $\tilde 1$}}
\put(340.0,150.00){\makebox(0.00,0.00){$+$}}
\put(350.00,150.00){\line(1,0){20.00}}
\put(410.00,150.00){\line(1,0){20.00}}
\qbezier(370.00,150.00)(390.00,170.00)(410.00,150.00)
\qbezier(370.00,150.00)(390.00,130.00)(410.00,150.00)
\put(360.0,140.00){\makebox(0.00,0.00){{\tiny $\tilde 1$}}}
\put(420.0,140.00){\makebox(0.00,0.00){\tiny $\tilde 1$}}
\put(390.0,170.00){\makebox(0.00,0.00){\tiny $\tilde 1$}}
\put(390.0,130.00){\makebox(0.00,0.00){\tiny $\tilde 2$}}
\put(440.0,150.00){\makebox(0.00,0.00){$+$}}
\put(450.00,150.00){\line(1,0){20.00}}
\put(510.00,150.00){\line(1,0){20.00}}
\qbezier(470.00,150.00)(490.00,170.00)(510.00,150.00)
\qbezier(470.00,150.00)(490.00,130.00)(510.00,150.00)
\put(460.0,140.00){\makebox(0.00,0.00){{\tiny $\tilde 1$}}}
\put(520.0,140.00){\makebox(0.00,0.00){\tiny $\tilde 1$}}
\put(490.0,170.00){\makebox(0.00,0.00){\tiny $\tilde 2$}}
\put(490.0,130.00){\makebox(0.00,0.00){\tiny $\tilde 3$}}
\put(540.0,150.00){\makebox(0.00,0.00){$+$}}
\put(550.00,150.00){\line(1,0){20.00}}
\put(610.00,150.00){\line(1,0){20.00}}
\qbezier(570.00,150.00)(590.00,170.00)(610.00,150.00)
\qbezier(570.00,150.00)(590.00,130.00)(610.00,150.00)
\put(560.0,140.00){\makebox(0.00,0.00){{\tiny $\tilde 1$}}}
\put(620.0,140.00){\makebox(0.00,0.00){\tiny $\tilde 1$}}
\put(590.0,170.00){\makebox(0.00,0.00){\tiny $\tilde 2$}}
\put(590.0,130.00){\makebox(0.00,0.00){\tiny $\tilde 4$}}
\put(640.0,150.00){\makebox(0.00,0.00){$+$}}
\put(650.00,150.00){\line(1,0){20.00}}
\put(710.00,150.00){\line(1,0){20.00}}
\qbezier(670.00,150.00)(690.00,170.00)(710.00,150.00)
\qbezier(670.00,150.00)(690.00,130.00)(710.00,150.00)
\put(660.0,140.00){\makebox(0.00,0.00){{\tiny $\tilde 1$}}}
\put(720.0,140.00){\makebox(0.00,0.00){\tiny $\tilde 1$}}
\put(690.0,170.00){\makebox(0.00,0.00){\tiny $\tilde 3$}}
\put(690.0,130.00){\makebox(0.00,0.00){\tiny $\tilde 4$}}
\put(740.0,150.00){\makebox(0.00,0.00){$+$}}
\put(250.00,70.00){\line(1,0){20.00}}
\put(310.00,70.00){\line(1,0){20.00}}
\qbezier(270.00,70.00)(290.00,90.00)(310.00,70.00)
\qbezier(270.00,70.00)(290.00,50.00)(310.00,70.00)
\put(260.0,60.00){\makebox(0.00,0.00){{\tiny $\tilde 1$}}}
\put(320.0,60.00){\makebox(0.00,0.00){\tiny $\tilde 1$}}
\put(290.0,90.00){\makebox(0.00,0.00){\tiny $\tilde 2$}}
\put(290.0,50.00){\makebox(0.00,0.00){\tiny $\tilde 2$}}
\put(340.0,70.00){\makebox(0.00,0.00){$+$}}
\put(350.00,70.00){\line(1,0){20.00}}
\put(410.00,70.00){\line(1,0){20.00}}
\qbezier(370.00,70.00)(390.00,90.00)(410.00,70.00)
\qbezier(370.00,70.00)(390.00,50.00)(410.00,70.00)
\put(360.0,60.00){\makebox(0.00,0.00){{\tiny $\tilde 1$}}}
\put(420.0,60.00){\makebox(0.00,0.00){\tiny $\tilde 1$}}
\put(390.0,90.00){\makebox(0.00,0.00){\tiny $\tilde 3$}}
\put(390.0,50.00){\makebox(0.00,0.00){\tiny $\tilde 3$}}
\put(440.0,70.00){\makebox(0.00,0.00){$+$}}
\put(450.00,70.700){\line(1,0){20.00}}
\put(510.00,70.00){\line(1,0){20.00}}
\qbezier(470.00,70.00)(490.00,90.00)(510.00,70.00)
\qbezier(470.00,70.00)(490.00,50.00)(510.00,70.00)
\put(460.0,60.00){\makebox(0.00,0.00){{\tiny $\tilde 1$}}}
\put(520.0,60.00){\makebox(0.00,0.00){\tiny $\tilde 1$}}
\put(490.0,90.00){\makebox(0.00,0.00){\tiny $\tilde 4$}}
\put(490.0,50.00){\makebox(0.00,0.00){\tiny $\tilde 4$}}
\end{picture}\bigskip \medskip \bigskip

\unitlength=0.680000pt 
\begin{picture}(237.92,115.00)(180.00,45.00)

\put(200.0,150.00){\makebox(0.00,0.00){$\Sigma (\tilde m_2^2) \quad = $}}

\put(250.00,150.00){\line(1,0){20.00}}
\put(310.00,150.00){\line(1,0){20.00}}
\qbezier(270.00,150.00)(290.00,170.00)(310.00,150.00)
\qbezier(270.00,150.00)(290.00,130.00)(310.00,150.00)
\put(260.0,140.00){\makebox(0.00,0.00){{\tiny $\tilde 2$}}}
\put(320.0,140.00){\makebox(0.00,0.00){\tiny $\tilde 2$}}
\put(290.0,170.00){\makebox(0.00,0.00){\tiny $\tilde 1$}}
\put(290.0,130.00){\makebox(0.00,0.00){\tiny $\tilde 1$}}
\put(340.0,150.00){\makebox(0.00,0.00){$+$}}
\put(350.00,150.00){\line(1,0){20.00}}
\put(410.00,150.00){\line(1,0){20.00}}
\qbezier(370.00,150.00)(390.00,170.00)(410.00,150.00)
\qbezier(370.00,150.00)(390.00,130.00)(410.00,150.00)
\put(360.0,140.00){\makebox(0.00,0.00){{\tiny $\tilde 2$}}}
\put(420.0,140.00){\makebox(0.00,0.00){\tiny $\tilde 2$}}
\put(390.0,170.00){\makebox(0.00,0.00){\tiny $\tilde 1$}}
\put(390.0,130.00){\makebox(0.00,0.00){\tiny $\tilde 2$}}
\put(440.0,150.00){\makebox(0.00,0.00){$+$}}
\put(450.00,150.00){\line(1,0){20.00}}
\put(510.00,150.00){\line(1,0){20.00}}
\qbezier(470.00,150.00)(490.00,170.00)(510.00,150.00)
\qbezier(470.00,150.00)(490.00,130.00)(510.00,150.00)
\put(460.0,140.00){\makebox(0.00,0.00){{\tiny $\tilde 2$}}}
\put(520.0,140.00){\makebox(0.00,0.00){\tiny $\tilde 2$}}
\put(490.0,170.00){\makebox(0.00,0.00){\tiny $\tilde 1$}}
\put(490.0,130.00){\makebox(0.00,0.00){\tiny $\tilde 3$}}
\put(540.0,150.00){\makebox(0.00,0.00){$+$}}
\put(550.00,150.00){\line(1,0){20.00}}
\put(610.00,150.00){\line(1,0){20.00}}
\qbezier(570.00,150.00)(590.00,170.00)(610.00,150.00)
\qbezier(570.00,150.00)(590.00,130.00)(610.00,150.00)
\put(560.0,140.00){\makebox(0.00,0.00){{\tiny $\tilde 2$}}}
\put(620.0,140.00){\makebox(0.00,0.00){\tiny $\tilde 2$}}
\put(590.0,170.00){\makebox(0.00,0.00){\tiny $\tilde 1$}}
\put(590.0,130.00){\makebox(0.00,0.00){\tiny $\tilde 4$}}
\put(640.0,150.00){\makebox(0.00,0.00){$+$}}
\put(650.00,150.00){\line(1,0){20.00}}
\put(710.00,150.00){\line(1,0){20.00}}
\qbezier(670.00,150.00)(690.00,170.00)(710.00,150.00)
\qbezier(670.00,150.00)(690.00,130.00)(710.00,150.00)
\put(660.0,140.00){\makebox(0.00,0.00){{\tiny $\tilde 2$}}}
\put(720.0,140.00){\makebox(0.00,0.00){\tiny $\tilde 2$}}
\put(690.0,170.00){\makebox(0.00,0.00){\tiny $\tilde 2$}}
\put(690.0,130.00){\makebox(0.00,0.00){\tiny $\tilde 3$}}
\put(740.0,150.00){\makebox(0.00,0.00){$+$}}
\put(250.00,70.00){\line(1,0){20.00}}
\put(310.00,70.00){\line(1,0){20.00}}
\qbezier(270.00,70.00)(290.00,90.00)(310.00,70.00)
\qbezier(270.00,70.00)(290.00,50.00)(310.00,70.00)
\put(260.0,60.00){\makebox(0.00,0.00){{\tiny $\tilde 2$}}}
\put(320.0,60.00){\makebox(0.00,0.00){\tiny $\tilde 2$}}
\put(290.0,90.00){\makebox(0.00,0.00){\tiny $\tilde 2$}}
\put(290.0,50.00){\makebox(0.00,0.00){\tiny $\tilde 4$}}
\put(340.0,70.00){\makebox(0.00,0.00){$+$}}
\put(350.00,70.00){\line(1,0){20.00}}
\put(410.00,70.00){\line(1,0){20.00}}
\qbezier(370.00,70.00)(390.00,90.00)(410.00,70.00)
\qbezier(370.00,70.00)(390.00,50.00)(410.00,70.00)
\put(360.0,60.00){\makebox(0.00,0.00){{\tiny $\tilde 2$}}}
\put(420.0,60.00){\makebox(0.00,0.00){\tiny $\tilde 2$}}
\put(390.0,90.00){\makebox(0.00,0.00){\tiny $\tilde 3$}}
\put(390.0,50.00){\makebox(0.00,0.00){\tiny $\tilde 4$}}
\put(440.0,70.00){\makebox(0.00,0.00){$+$}}
\put(450.00,70.700){\line(1,0){20.00}}
\put(510.00,70.00){\line(1,0){20.00}}
\qbezier(470.00,70.00)(490.00,90.00)(510.00,70.00)
\qbezier(470.00,70.00)(490.00,50.00)(510.00,70.00)
\put(460.0,60.00){\makebox(0.00,0.00){{\tiny $\tilde 2$}}}
\put(520.0,60.00){\makebox(0.00,0.00){\tiny $\tilde 2$}}
\put(490.0,90.00){\makebox(0.00,0.00){\tiny $\tilde 2$}}
\put(490.0,50.00){\makebox(0.00,0.00){\tiny $\tilde 2$}}
\put(540.0,70.00){\makebox(0.00,0.00){$+$}}
\put(550.00,70.00){\line(1,0){20.00}}
\put(610.00,70.00){\line(1,0){20.00}}
\qbezier(570.00,70.00)(590.00,90.00)(610.00,70.00)
\qbezier(570.00,70.00)(590.00,50.00)(610.00,70.00)
\put(560.0,60.00){\makebox(0.00,0.00){{\tiny $\tilde 2$}}}
\put(620.0,60.00){\makebox(0.00,0.00){\tiny $\tilde 2$}}
\put(590.0,90.00){\makebox(0.00,0.00){\tiny $\tilde 3$}}
\put(590.0,50.00){\makebox(0.00,0.00){\tiny $\tilde 3$}}
\put(640.0,70.00){\makebox(0.00,0.00){$+$}}
\put(650.00,70.00){\line(1,0){20.00}}
\put(710.00,70.00){\line(1,0){20.00}}
\qbezier(670.00,70.00)(690.00,90.00)(710.00,70.00)
\qbezier(670.00,70.00)(690.00,50.00)(710.00,70.00)
\put(660.0,60.00){\makebox(0.00,0.00){{\tiny $\tilde 2$}}}
\put(720.0,60.00){\makebox(0.00,0.00){\tiny $\tilde 2$}}
\put(690.0,90.00){\makebox(0.00,0.00){\tiny $\tilde 4$}}
\put(690.0,50.00){\makebox(0.00,0.00){\tiny $\tilde 4$}}
\end{picture}\newpage

\bigskip \medskip \unitlength=0.680000pt 
\begin{picture}(237.92,120.00)(180.00,50.00)

\put(200.0,150.00){\makebox(0.00,0.00){$\Sigma (\tilde m_3^2) \quad = $}}

\put(250.00,150.00){\line(1,0){20.00}}
\put(310.00,150.00){\line(1,0){20.00}}
\qbezier(270.00,150.00)(290.00,170.00)(310.00,150.00)
\qbezier(270.00,150.00)(290.00,130.00)(310.00,150.00)
\put(260.0,140.00){\makebox(0.00,0.00){{\tiny $\tilde 3$}}}
\put(320.0,140.00){\makebox(0.00,0.00){\tiny $\tilde 3$}}
\put(290.0,170.00){\makebox(0.00,0.00){\tiny $\tilde 1$}}
\put(290.0,130.00){\makebox(0.00,0.00){\tiny $\tilde 2$}}
\put(340.0,150.00){\makebox(0.00,0.00){$+$}}
\put(350.00,150.00){\line(1,0){20.00}}
\put(410.00,150.00){\line(1,0){20.00}}
\qbezier(370.00,150.00)(390.00,170.00)(410.00,150.00)
\qbezier(370.00,150.00)(390.00,130.00)(410.00,150.00)
\put(360.0,140.00){\makebox(0.00,0.00){{\tiny $\tilde 3$}}}
\put(420.0,140.00){\makebox(0.00,0.00){\tiny $\tilde 3$}}
\put(390.0,170.00){\makebox(0.00,0.00){\tiny $\tilde 1$}}
\put(390.0,130.00){\makebox(0.00,0.00){\tiny $\tilde 3$}}
\put(440.0,150.00){\makebox(0.00,0.00){$+$}}
\put(450.00,150.00){\line(1,0){20.00}}
\put(510.00,150.00){\line(1,0){20.00}}
\qbezier(470.00,150.00)(490.00,170.00)(510.00,150.00)
\qbezier(470.00,150.00)(490.00,130.00)(510.00,150.00)
\put(460.0,140.00){\makebox(0.00,0.00){{\tiny $\tilde 3$}}}
\put(520.0,140.00){\makebox(0.00,0.00){\tiny $\tilde 3$}}
\put(490.0,170.00){\makebox(0.00,0.00){\tiny $\tilde 1$}}
\put(490.0,130.00){\makebox(0.00,0.00){\tiny $\tilde 4$}}
\put(540.0,150.00){\makebox(0.00,0.00){$+$}}
\put(550.00,150.00){\line(1,0){20.00}}
\put(610.00,150.00){\line(1,0){20.00}}
\qbezier(570.00,150.00)(590.00,170.00)(610.00,150.00)
\qbezier(570.00,150.00)(590.00,130.00)(610.00,150.00)
\put(560.0,140.00){\makebox(0.00,0.00){{\tiny $\tilde 3$}}}
\put(620.0,140.00){\makebox(0.00,0.00){\tiny $\tilde 3$}}
\put(590.0,170.00){\makebox(0.00,0.00){\tiny $\tilde 2$}}
\put(590.0,130.00){\makebox(0.00,0.00){\tiny $\tilde 3$}}
\put(640.0,150.00){\makebox(0.00,0.00){$+$}}
\put(650.00,150.00){\line(1,0){20.00}}
\put(710.00,150.00){\line(1,0){20.00}}
\qbezier(670.00,150.00)(690.00,170.00)(710.00,150.00)
\qbezier(670.00,150.00)(690.00,130.00)(710.00,150.00)
\put(660.0,140.00){\makebox(0.00,0.00){{\tiny $\tilde 3$}}}
\put(720.0,140.00){\makebox(0.00,0.00){\tiny $\tilde 3$}}
\put(690.0,170.00){\makebox(0.00,0.00){\tiny $\tilde 2$}}
\put(690.0,130.00){\makebox(0.00,0.00){\tiny $\tilde 4$}}
\put(740.0,150.00){\makebox(0.00,0.00){$+$}}
\put(250.00,70.00){\line(1,0){20.00}}
\put(310.00,70.00){\line(1,0){20.00}}
\qbezier(270.00,70.00)(290.00,90.00)(310.00,70.00)
\qbezier(270.00,70.00)(290.00,50.00)(310.00,70.00)
\put(260.0,60.00){\makebox(0.00,0.00){{\tiny $\tilde 3$}}}
\put(320.0,60.00){\makebox(0.00,0.00){\tiny $\tilde 3$}}
\put(290.0,90.00){\makebox(0.00,0.00){\tiny $\tilde 3$}}
\put(290.0,50.00){\makebox(0.00,0.00){\tiny $\tilde 4$}}
\put(340.0,70.00){\makebox(0.00,0.00){$+$}}
\put(350.00,70.00){\line(1,0){20.00}}
\put(410.00,70.00){\line(1,0){20.00}}
\qbezier(370.00,70.00)(390.00,90.00)(410.00,70.00)
\qbezier(370.00,70.00)(390.00,50.00)(410.00,70.00)
\put(360.0,60.00){\makebox(0.00,0.00){{\tiny $\tilde 3$}}}
\put(420.0,60.00){\makebox(0.00,0.00){\tiny $\tilde 3$}}
\put(390.0,90.00){\makebox(0.00,0.00){\tiny $\tilde 2$}}
\put(390.0,50.00){\makebox(0.00,0.00){\tiny $\tilde 2$}}
\put(440.0,70.00){\makebox(0.00,0.00){$+$}}
\put(450.00,70.700){\line(1,0){20.00}}
\put(510.00,70.00){\line(1,0){20.00}}
\qbezier(470.00,70.00)(490.00,90.00)(510.00,70.00)
\qbezier(470.00,70.00)(490.00,50.00)(510.00,70.00)
\put(460.0,60.00){\makebox(0.00,0.00){{\tiny $\tilde 3$}}}
\put(520.0,60.00){\makebox(0.00,0.00){\tiny $\tilde 3$}}
\put(490.0,90.00){\makebox(0.00,0.00){\tiny $\tilde 3$}}
\put(490.0,50.00){\makebox(0.00,0.00){\tiny $\tilde 3$}}
\put(540.0,70.00){\makebox(0.00,0.00){$+$}}
\put(550.00,70.00){\line(1,0){20.00}}
\put(610.00,70.00){\line(1,0){20.00}}
\qbezier(570.00,70.00)(590.00,90.00)(610.00,70.00)
\qbezier(570.00,70.00)(590.00,50.00)(610.00,70.00)
\put(560.0,60.00){\makebox(0.00,0.00){{\tiny $\tilde 3$}}}
\put(620.0,60.00){\makebox(0.00,0.00){\tiny $\tilde 3$}}
\put(590.0,90.00){\makebox(0.00,0.00){\tiny $\tilde 4$}}
\put(590.0,50.00){\makebox(0.00,0.00){\tiny $\tilde 4$}}
\end{picture}\bigskip \medskip \bigskip

\unitlength=0.680000pt 
\begin{picture}(237.92,115.00)(180.00,45.00)

\put(200.0,150.00){\makebox(0.00,0.00){$\Sigma (\tilde m_4^2) \quad = $}}

\put(250.00,150.00){\line(1,0){20.00}}
\put(310.00,150.00){\line(1,0){20.00}}
\qbezier(270.00,150.00)(290.00,170.00)(310.00,150.00)
\qbezier(270.00,150.00)(290.00,130.00)(310.00,150.00)
\put(260.0,140.00){\makebox(0.00,0.00){{\tiny $\tilde 4$}}}
\put(320.0,140.00){\makebox(0.00,0.00){\tiny $\tilde 4$}}
\put(290.0,170.00){\makebox(0.00,0.00){\tiny $\tilde 1$}}
\put(290.0,130.00){\makebox(0.00,0.00){\tiny $\tilde 2$}}
\put(340.0,150.00){\makebox(0.00,0.00){$+$}}
\put(350.00,150.00){\line(1,0){20.00}}
\put(410.00,150.00){\line(1,0){20.00}}
\qbezier(370.00,150.00)(390.00,170.00)(410.00,150.00)
\qbezier(370.00,150.00)(390.00,130.00)(410.00,150.00)
\put(360.0,140.00){\makebox(0.00,0.00){{\tiny $\tilde 4$}}}
\put(420.0,140.00){\makebox(0.00,0.00){\tiny $\tilde 4$}}
\put(390.0,170.00){\makebox(0.00,0.00){\tiny $\tilde 1$}}
\put(390.0,130.00){\makebox(0.00,0.00){\tiny $\tilde 3$}}
\put(440.0,150.00){\makebox(0.00,0.00){$+$}}
\put(450.00,150.00){\line(1,0){20.00}}
\put(510.00,150.00){\line(1,0){20.00}}
\qbezier(470.00,150.00)(490.00,170.00)(510.00,150.00)
\qbezier(470.00,150.00)(490.00,130.00)(510.00,150.00)
\put(460.0,140.00){\makebox(0.00,0.00){{\tiny $\tilde 4$}}}
\put(520.0,140.00){\makebox(0.00,0.00){\tiny $\tilde 4$}}
\put(490.0,170.00){\makebox(0.00,0.00){\tiny $\tilde 1$}}
\put(490.0,130.00){\makebox(0.00,0.00){\tiny $\tilde 4$}}
\put(540.0,150.00){\makebox(0.00,0.00){$+$}}
\put(550.00,150.00){\line(1,0){20.00}}
\put(610.00,150.00){\line(1,0){20.00}}
\qbezier(570.00,150.00)(590.00,170.00)(610.00,150.00)
\qbezier(570.00,150.00)(590.00,130.00)(610.00,150.00)
\put(560.0,140.00){\makebox(0.00,0.00){{\tiny $\tilde 4$}}}
\put(620.0,140.00){\makebox(0.00,0.00){\tiny $\tilde 4$}}
\put(590.0,170.00){\makebox(0.00,0.00){\tiny $\tilde 2$}}
\put(590.0,130.00){\makebox(0.00,0.00){\tiny $\tilde 3$}}
\put(640.0,150.00){\makebox(0.00,0.00){$+$}}
\put(650.00,150.00){\line(1,0){20.00}}
\put(710.00,150.00){\line(1,0){20.00}}
\qbezier(670.00,150.00)(690.00,170.00)(710.00,150.00)
\qbezier(670.00,150.00)(690.00,130.00)(710.00,150.00)
\put(660.0,140.00){\makebox(0.00,0.00){{\tiny $\tilde 4$}}}
\put(720.0,140.00){\makebox(0.00,0.00){\tiny $\tilde 4$}}
\put(690.0,170.00){\makebox(0.00,0.00){\tiny $\tilde 2$}}
\put(690.0,130.00){\makebox(0.00,0.00){\tiny $\tilde 4$}}
\put(740.0,150.00){\makebox(0.00,0.00){$+$}}
\put(250.00,70.00){\line(1,0){20.00}}
\put(310.00,70.00){\line(1,0){20.00}}
\qbezier(270.00,70.00)(290.00,90.00)(310.00,70.00)
\qbezier(270.00,70.00)(290.00,50.00)(310.00,70.00)
\put(260.0,60.00){\makebox(0.00,0.00){{\tiny $\tilde 4$}}}
\put(320.0,60.00){\makebox(0.00,0.00){\tiny $\tilde 4$}}
\put(290.0,90.00){\makebox(0.00,0.00){\tiny $\tilde 3$}}
\put(290.0,50.00){\makebox(0.00,0.00){\tiny $\tilde 4$}}
\put(340.0,70.00){\makebox(0.00,0.00){$+$}}
\put(350.00,70.00){\line(1,0){20.00}}
\put(410.00,70.00){\line(1,0){20.00}}
\qbezier(370.00,70.00)(390.00,90.00)(410.00,70.00)
\qbezier(370.00,70.00)(390.00,50.00)(410.00,70.00)
\put(360.0,60.00){\makebox(0.00,0.00){{\tiny $\tilde 4$}}}
\put(420.0,60.00){\makebox(0.00,0.00){\tiny $\tilde 4$}}
\put(390.0,90.00){\makebox(0.00,0.00){\tiny $\tilde 2$}}
\put(390.0,50.00){\makebox(0.00,0.00){\tiny $\tilde 2$}}
\put(440.0,70.00){\makebox(0.00,0.00){$+$}}
\put(450.00,70.700){\line(1,0){20.00}}
\put(510.00,70.00){\line(1,0){20.00}}
\qbezier(470.00,70.00)(490.00,90.00)(510.00,70.00)
\qbezier(470.00,70.00)(490.00,50.00)(510.00,70.00)
\put(460.0,60.00){\makebox(0.00,0.00){{\tiny $\tilde 4$}}}
\put(520.0,60.00){\makebox(0.00,0.00){\tiny $\tilde 4$}}
\put(490.0,90.00){\makebox(0.00,0.00){\tiny $\tilde 3$}}
\put(490.0,50.00){\makebox(0.00,0.00){\tiny $\tilde 3$}}
\put(540.0,70.00){\makebox(0.00,0.00){$+$}}
\put(550.00,70.00){\line(1,0){20.00}}
\put(610.00,70.00){\line(1,0){20.00}}
\qbezier(570.00,70.00)(590.00,90.00)(610.00,70.00)
\qbezier(570.00,70.00)(590.00,50.00)(610.00,70.00)
\put(560.0,60.00){\makebox(0.00,0.00){{\tiny $\tilde 4$}}}
\put(620.0,60.00){\makebox(0.00,0.00){\tiny $\tilde 4$}}
\put(590.0,90.00){\makebox(0.00,0.00){\tiny $\tilde 4$}}
\put(590.0,50.00){\makebox(0.00,0.00){\tiny $\tilde 4$}}
\end{picture}\medskip

\noindent From this it follows that the classical mass ratios are not
conserved in the quantum field theory 
\begin{equation}
\frac{\delta m_{1}^{2}}{m_{1}^{2}}=54045.1\ldots \quad \frac{\delta m_{2}^{2}%
}{m_{2}^{2}}=68239.3\ldots \quad \frac{\delta m_{3}^{2}}{m_{3}^{2}}%
=11488.2\ldots \quad \frac{\delta m_{4}^{2}}{m_{4}^{2}}=2914.28\ldots
\end{equation}
Hence we have the same type of behaviour under renormalisation as in the $%
H_{3}$-ATFT obtained from the reduction of the $D_{6}$-ATFT.

\section{Conclusions}

With regard to previously studied ATFT, the embedding of
non-crystallographic into crystallographic Coxeter groups leads to an
explanation for the fact that in some theories the masses can be organised
into pairs such that one mass differs from the other only by a factor of $%
\phi $. This also holds for the higher charges.

We showed that it is possible to construct ATFT related to
non-crystallographic Coxeter groups despite the fact that there is no Lie
algebra associated to them. The construction is possible since one may
exploit the embedding of non-crystallographic into crystallographic Coxeter
groups, making use of the fact that the latter do possess a Lie algebraic
structure and thus preserving integrability. Unlike the folding from simply
laced Lie algebras to non-simply laced Lie algebras \cite{DIO}, the
resulting theories we obtain here are not equivalent to a direct formulation
of the theory in terms of non-crystallographic Coxeter groups. In this
context the reduction procedure is vital for consistency and not merely an
additional structure. It is of course possible to write down a Lagrangian of
the form (\ref{L2}) involving directly roots of $\tilde{\Delta}$, but it
remains to be seen whether such theories are consistent and especially if
they are classically integrable.

With regard to the quantum field theory, our computations showed for the $%
H_{3},H_{4}$-ATFT that the masses of these new theories do not renormalise
with an overall factor, i.e. $\delta m_{k}^{2}/m_{k}^{2}$ is not a universal
constant for all particle types $k$, preventing that the classical mass
ratios can be maintained in the quantum field theory. Remarkably this was
true for ATFT related to simply laced Lie algebras, which allowed for a
relatively straightforward construction of the scattering matrices \cite%
{BCDS}. For ATFT related to non-simply laced Lie algebras this was found no
longer to be true, such that different types of scenarios had to be devised.
One is to have floating masses such that depending on the coupling constant
the masses flow from one Lie algebra in the weak limit to its Langlands dual
in the strong coupling limit \cite{Del3,KW,WW,CDS,PD,PK,TO,q2}. The other
alternative proposal was to introduce additional Fermions into the model 
\cite{Del1,Del2,GPZ,DVF}, which compensate for the unequal mass shifts. From
our analysis it is clear that the construction of a consistent quantum field
theory for the proposed $H_{3},H_{4}$-theories has to be modelled along the
line of the construction of theories related to non-simply laced algebras
due to unequal mass renormalisations for each individual particle. It
remains to be seen in future work, which of the prescriptions will be
successful in this context.

As we showed, the behaviour under renormalisation is different for the $%
H_{2} $-ATFT, where the classical mass ratios remain preserved up to first
order perturbation theory. Despite this, it is not immediately obvious how
to write down a scattering matrix to all orders in perturbation theory.

Our detailed analysis of the embedding of non-crystallographic into
crystallographic Coxeter groups allows one to apply the aforementioned
reduction method to a wide range of application in physics, chemistry and
biology, where Coxeter groups play a role. For instance in our forthcoming
publication \cite{FK} we will apply this method to the generalized
Calogero-Moser models.\bigskip

\noindent \textbf{Acknowledgments}. C.K. is financially supported by a
University Research Fellowship of the Royal Society.

\newpage

\begin{center}
\noindent \textbf{\large Appendix}
\end{center}

\appendix

\section{The orbits of $H_{3}$ and $D_{6}$}

Successive action of $\sigma =$ $\sigma _{1}\sigma _{4}\sigma _{3}\sigma
_{6}\sigma _{2}\sigma _{5}$ and $\tilde{\sigma}=$ $\tilde{\sigma}_{1}\tilde{%
\sigma}_{3}\tilde{\sigma}_{2}$ yields

\begin{center}
\begin{tabular}{||l||c|c|c|c||}
\hline
& $\Omega _{1}$ & $\omega (\Omega _{1})=\tilde{\Omega}_{1}$ & $\Omega _{4}$
& $\omega (\Omega _{4})=\phi \tilde{\Omega}_{1}$ \\ \hline\hline
$\sigma ^{0}$ & ${{\alpha }_{1}}$ & ${{\tilde{\alpha}}_{1}}$ & ${{\alpha }%
_{4}}$ & $\phi {{\tilde{\alpha}}_{1}}$ \\ \hline
$\sigma ^{1}$ & ${{\alpha }_{2}}+{{\alpha }_{6}}$ & ${{\tilde{\alpha}}_{2}}%
+\phi \,{{\tilde{\alpha}}_{3}}$ & ${{\alpha }_{3}}+{{\alpha }_{5}}+{{\alpha }%
_{6}}$ & $\phi {{\tilde{\alpha}}_{2}}+\phi ^{2}\,{{\tilde{\alpha}}_{3}}$ \\ 
\hline
$\sigma ^{2}$ & ${{\alpha }_{3}}+{{\alpha }_{4}}+{{\alpha }_{5}}$ & $\phi {{%
\tilde{\alpha}}_{1}}+\phi {{\tilde{\alpha}}_{2}}+{{\tilde{\alpha}}_{3}}$ & ${%
{\alpha }_{1}}+{{\alpha }_{2}}+{{\alpha }_{4}}+{{\alpha }_{5}}+{{\alpha }_{6}%
}$ & $\phi ^{2}{{\tilde{\alpha}}_{1}}+\phi ^{2}{{\tilde{\alpha}}_{2}}+\phi {{%
\tilde{\alpha}}_{3}}$ \\ \hline
$\sigma ^{3}$ & ${{\alpha }_{5}}+{{\alpha }_{6}}$ & $\phi \left( {{\tilde{%
\alpha}}_{2}}+{{\tilde{\alpha}}_{3}}\right) $ & ${{{\alpha }_{2}}}+{{{\alpha 
}_{3}}}+{{{\alpha }_{5}}}+{{{\alpha }_{6}}}$ & $\phi ^{2}\left( {{\tilde{%
\alpha}}_{2}}+{{\tilde{\alpha}}_{3}}\right) $ \\ \hline
$\sigma ^{4}$ & ${{\alpha }_{1}}+{{\alpha }_{2}}$ & ${{\tilde{\alpha}}_{1}}+{%
{\tilde{\alpha}}_{2}}$ & ${{\alpha }_{4}}+{{\alpha }_{5}}$ & $\phi ({{\tilde{%
\alpha}}_{1}}+{{\tilde{\alpha}}_{2})}$ \\ \hline
$\sigma ^{5}$ & $-{{\alpha }_{1}}$ & $-{{\tilde{\alpha}}_{1}}$ & $-{{\alpha }%
_{4}}$ & $-\phi {{\tilde{\alpha}}_{1}}$ \\ \hline
\end{tabular}

\begin{tabular}{||l||c|c|c||}
\hline
& $\Omega _{2}$ & $\omega (\Omega _{2})=\tilde{\Omega}_{2}$ & $\Omega _{5}$
\\ \hline\hline
$\sigma ^{0}$ & $-{{\alpha }_{2}}$ & $-{{\tilde{\alpha}}_{2}}$ & $-{{\alpha }%
_{5}}$ \\ \hline
$\sigma ^{1}$ & ${{\alpha }_{1}}+{{\alpha }_{2}}+{{\alpha }_{6}}$ & ${{%
\tilde{\alpha}}_{1}}+{{\tilde{\alpha}}_{2}}+\phi {{\tilde{\alpha}}_{3}}$ & ${%
{\alpha }_{3}}+{{\alpha }_{4}}+{{\alpha }_{5}}+{{\alpha }_{6}}$ \\ \hline
$\sigma ^{2}$ & ${{\alpha }_{2}}+{{\alpha }_{3}}+{{\alpha }_{4}}+{{\alpha }%
_{5}}+{{\alpha }_{6}}$ & $\phi {{\tilde{\alpha}}_{1}}+\phi {{\tilde{\alpha}}%
_{2}}+\phi ^{2}{{\tilde{\alpha}}_{3}}$ & ${{{{\alpha }_{1}}+\alpha }_{2}}+{{%
\alpha }_{3}}+{{\alpha }_{4}}+2{{\alpha }_{5}}+2{{\alpha }_{6}}$ \\ \hline
$\sigma ^{3}$ & ${{\alpha }_{3}}+{{\alpha }_{4}}+2{{\alpha }_{5}}+{{\alpha }%
_{6}}$ & $\phi {{\tilde{\alpha}}_{1}}+2\phi {{\tilde{\alpha}}_{2}}+\phi ^{2}{%
{\tilde{\alpha}}_{3}}$ & ${{{{\alpha }_{1}}+2\alpha }_{2}}+{{\alpha }_{3}}+{{%
\alpha }_{4}}+2{{\alpha }_{5}}+2{{\alpha }_{6}}$ \\ \hline
$\sigma ^{4}$ & ${{\alpha }_{1}}+{{\alpha }_{2}}+{{\alpha }_{5}}+{{\alpha }%
_{6}}$ & ${{\tilde{\alpha}}_{1}}+\phi ^{2}{{\tilde{\alpha}}_{2}}+\phi {{%
\tilde{\alpha}}_{3}}$ & ${{\alpha }_{2}}+{{\alpha }_{3}}+{{\alpha }_{4}}+2{{%
\alpha }_{5}}+{{\alpha }_{6}}$ \\ \hline
$\sigma ^{5}$ & ${{\alpha }_{2}}$ & ${{\tilde{\alpha}}_{2}}$ & ${{\alpha }%
_{5}}$ \\ \hline\hline
& $\Omega _{3}$ & $\omega (\Omega _{3})=\tilde{\Omega}_{3}$ & $\Omega _{6}$
\\ \hline\hline
$\sigma ^{0}$ & ${{\alpha }_{3}}$ & ${{\tilde{\alpha}}_{3}}$ & ${{\alpha }%
_{6}}$ \\ \hline
$\sigma ^{1}$ & ${{\alpha }_{4}}+{{\alpha }_{5}}+{{\alpha }_{6}}$ & $\phi ({{%
\tilde{\alpha}}_{1}}+{{\tilde{\alpha}}_{2}}+\,{{\tilde{\alpha}}_{3})}$ & ${{%
\alpha }_{1}}+{{\alpha }_{2}+{\alpha }_{3}}+{{\alpha }_{4}}+{{\alpha }_{5}}+{%
{\alpha }_{6}}$ \\ \hline
$\sigma ^{2}$ & ${{\alpha }_{1}}+{{\alpha }_{2}}+{{\alpha }_{3}}+{{\alpha }%
_{5}}+{{\alpha }_{6}}$ & ${{\tilde{\alpha}}_{1}}+\phi ^{2}{{\tilde{\alpha}}%
_{2}}+\phi ^{2}{{\tilde{\alpha}}_{3}}$ & ${{\alpha }_{2}+{\alpha }_{3}}+{{%
\alpha }_{4}}+2{{\alpha }_{5}}+2{{\alpha }_{6}}$ \\ \hline
$\sigma ^{3}$ & ${{\alpha }_{2}}+{{\alpha }_{4}}+{{\alpha }_{5}}+{{\alpha }%
_{6}}$ & $\phi ({{\tilde{\alpha}}_{1}}+\phi {{\tilde{\alpha}}_{2}}+\,{{%
\tilde{\alpha}}_{3})}$ & ${{{{\alpha }_{1}}+\alpha }_{2}}+{{\alpha }_{3}}+{{%
\alpha }_{4}}+2{{\alpha }_{5}}+{{\alpha }_{6}}$ \\ \hline
$\sigma ^{4}$ & ${{\alpha }_{3}}+{{\alpha }_{5}}$ & $\phi {{\tilde{\alpha}}%
_{2}}+{{\tilde{\alpha}}_{3}}$ & ${{\alpha }_{2}}+{{\alpha }_{5}}+{{\alpha }%
_{6}}$ \\ \hline
$\sigma ^{5}$ & $-{{\alpha }_{3}}$ & $-{{\tilde{\alpha}}_{3}}$ & $-{{\alpha }%
_{6}}$ \\ \hline
\end{tabular}
\end{center}

\noindent We did not write here the additional $\tilde{\sigma}^{i}$ in the
first column. The identities\ $\omega (\Omega _{4})=\phi \tilde{\Omega}_{1}$%
, $\omega (\Omega _{4})=\phi \tilde{\Omega}_{1}$and $\omega (\Omega
_{6})=\phi \tilde{\Omega}_{3}$ follow upon using (\ref{tau}).

\newpage

\section{The orbits of $H_{4}$ and $E_{8}$}

Successive action of $\sigma =$ $\sigma _{1}\sigma _{5}\sigma _{3}\sigma
_{7}\sigma _{2}\sigma _{6}\sigma _{4}\sigma _{8}$ and $\tilde{\sigma}=$ $%
\tilde{\sigma}_{1}\tilde{\sigma}_{3}\tilde{\sigma}_{2}\tilde{\sigma}_{4}$
yields

\begin{center}
\begin{tabular}{||l||l|l|l||}
\hline
& $\Omega _{1}$ & $\omega (\Omega _{1})=\tilde{\Omega}_{1}$ & $\Omega _{5}$
\\ \hline\hline
$\sigma ^{0}$ & $\alpha _{1}$ & $\tilde{\alpha}_{1}$ & $\alpha _{5}$ \\ 
\hline
$\sigma ^{1}$ & $\alpha _{2}+\alpha _{3}$ & $\tilde{\alpha}_{2}+\tilde{\alpha%
}_{3}$ & $\alpha _{6}+\alpha _{7}$ \\ \hline
$\sigma ^{2}$ & $\alpha _{7}+\alpha _{8}$ & $\phi (\tilde{\alpha}_{3}+\tilde{%
\alpha}_{4})$ & $\alpha _{3}+\alpha _{4}+\alpha _{7}+\alpha _{8}$ \\ \hline
$\sigma ^{3}$ & $\alpha _{4}+\alpha _{5}+\alpha _{7}+\alpha _{8}$ & $\phi (%
\tilde{\alpha}_{1}+\tilde{\alpha}_{2}+\tilde{\alpha}_{3})+\tilde{\alpha}_{4}$
& $%
\begin{array}{l}
\alpha _{1}+\alpha _{2}+\alpha _{3}+\alpha _{5} \\ 
+\alpha _{6}+\alpha _{7}+\alpha _{8}%
\end{array}
$ \\ \hline
$\sigma ^{4}$ & $\alpha _{3}+\alpha _{6}+\alpha _{7}+\alpha _{8}$ & $\phi 
\tilde{\alpha}_{2}+\phi ^{2}\tilde{\alpha}_{3}+\phi \tilde{\alpha}_{4}$ & $%
\begin{array}{l}
\alpha _{2}+\alpha _{3}+\alpha _{4}+\alpha _{6} \\ 
+2\alpha _{7}+\alpha _{8}%
\end{array}
$ \\ \hline
$\sigma ^{5}$ & $\alpha _{1}+\alpha _{2}+\alpha _{3}+\alpha _{4}+\alpha
_{7}+\alpha _{8}$ & $\tilde{\alpha}_{1}+\tilde{\alpha}_{2}+\phi ^{2}(\tilde{%
\alpha}_{3}+\tilde{\alpha}_{4})$ & $%
\begin{array}{l}
\alpha _{3}+\alpha _{4}+\alpha _{5}+\alpha _{6} \\ 
+2\alpha _{7}+\alpha _{8}%
\end{array}
$ \\ \hline
$\sigma ^{6}$ & $\alpha _{2}+\alpha _{3}+\alpha _{5}+\alpha _{6}+\alpha
_{7}+\alpha _{8}$ & $\phi (\tilde{\alpha}_{1}+\tilde{\alpha}_{4})+\phi ^{2}(%
\tilde{\alpha}_{2}+\tilde{\alpha}_{3})$ & $%
\begin{array}{l}
\alpha _{1}+\alpha _{2}+\alpha _{3}+\alpha _{4} \\ 
+\alpha _{5}+2\alpha _{6}+2\alpha _{7}+\alpha _{8}%
\end{array}
$ \\ \hline
$\sigma ^{7}$ & $\alpha _{4}+\alpha _{6}+2\alpha _{7}+\alpha _{8}$ & $\phi 
\tilde{\alpha}_{2}+2\phi \tilde{\alpha}_{3}+\phi ^{2}\tilde{\alpha}_{4}$ & $%
\begin{array}{l}
\alpha _{2}+2\alpha _{3}+\alpha _{4}+\alpha _{5} \\ 
+\alpha _{6}+2\alpha _{7}+2\alpha _{8}%
\end{array}
$ \\ \hline
$\sigma ^{8}$ & $\alpha _{3}+\alpha _{4}+\alpha _{5}+\alpha _{6}+\alpha
_{7}+\alpha _{8}$ & $\phi (\tilde{\alpha}_{1}+\tilde{\alpha}_{2})+\phi ^{2}(%
\tilde{\alpha}_{3}+\tilde{\alpha}_{4})$ & $%
\begin{array}{l}
\alpha _{1}+\alpha _{2}+\alpha _{3}+\alpha _{4} \\ 
+\alpha _{5}+\alpha _{6}+2\alpha _{7}+2\alpha _{8}%
\end{array}
$ \\ \hline
$\sigma ^{9}$ & $\alpha _{1}+\alpha _{2}+\alpha _{3}+\alpha _{6}+\alpha
_{7}+\alpha _{8}$ & $\tilde{\alpha}_{1}+\phi \tilde{\alpha}_{4}+\phi ^{2}(%
\tilde{\alpha}_{2}+\tilde{\alpha}_{3})$ & $%
\begin{array}{l}
\alpha _{2}+\alpha _{3}+\alpha _{4}+\alpha _{5} \\ 
+2\alpha _{6}+2\alpha _{7}+\alpha _{8}%
\end{array}
$ \\ \hline
$\sigma ^{10}$ & $\alpha _{2}+\alpha _{3}+\alpha _{4}+\alpha _{6}+\alpha
_{7}+\alpha _{8}$ & $\tilde{\alpha}_{2}+\phi ^{2}(\tilde{\alpha}_{3}+\tilde{%
\alpha}_{4})$ & $%
\begin{array}{l}
\alpha _{3}+\alpha _{4}+\alpha _{6}+2\alpha _{7} \\ 
+2\alpha _{8}%
\end{array}
$ \\ \hline
$\sigma ^{11}$ & $\alpha _{5}+\alpha _{6}+\alpha _{7}+\alpha _{8}$ & $\phi (%
\tilde{\alpha}_{1}+\tilde{\alpha}_{2}+\tilde{\alpha}_{3}+\tilde{\alpha}_{4})$
& $%
\begin{array}{l}
\alpha _{1}+\alpha _{2}+\alpha _{3}+\alpha _{4} \\ 
+\alpha _{5}+\alpha _{6}+\alpha _{7}+\alpha _{8}%
\end{array}
$ \\ \hline
$\sigma ^{12}$ & $\alpha _{4}+\alpha _{6}+\alpha _{7}$ & $\phi (\tilde{\alpha%
}_{2}+\tilde{\alpha}_{3})+\tilde{\alpha}_{4}$ & $%
\begin{array}{l}
\alpha _{2}+\alpha _{3}+\alpha _{6}+\alpha _{7} \\ 
+\alpha _{8}%
\end{array}
$ \\ \hline
$\sigma ^{13}$ & $\alpha _{3}+\alpha _{8}$ & $\tilde{\alpha}_{3}+\phi \tilde{%
\alpha}_{4}$ & $\alpha _{4}+\alpha _{7}+\alpha _{8}$ \\ \hline
$\sigma ^{14}$ & $\alpha _{1}+\alpha _{2}$ & $\tilde{\alpha}_{1}+\tilde{%
\alpha}_{2}$ & $\alpha _{5}+\alpha _{6}$ \\ \hline
$\sigma ^{15}$ & $-\alpha _{1}$ & $-\tilde{\alpha}_{1}$ & $-\alpha _{5}$ \\ 
\hline
\end{tabular}

\newpage

\begin{tabular}{||l||l|l|l||}
\hline
& $\Omega _{2}$ & $\omega (\Omega _{2})=\tilde{\Omega}_{2}$ & $\Omega _{6}$
\\ \hline
$\sigma ^{0}$ & $-\alpha _{2}$ & $\tilde{\alpha}_{2}$ & $-\alpha _{6}$ \\ 
\hline
$\sigma ^{1}$ & $\alpha _{1}+\alpha _{2}+\alpha _{3}$ & $\tilde{\alpha}_{1}+%
\tilde{\alpha}_{2}+\tilde{\alpha}_{3}$ & $\alpha _{5}+\alpha _{6}+\alpha
_{7} $ \\ \hline
$\sigma ^{2}$ & $\alpha _{2}+\alpha _{3}+\alpha _{7}+\alpha _{8}$ & $\tilde{%
\alpha}_{2}+${\small $\phi $}$^{2}\,\tilde{\alpha}_{3}+${\small $\phi $}$%
\tilde{\alpha}_{4}$ & $\alpha _{3}+\alpha _{4}+\alpha _{6}+2\,\alpha
_{7}+\alpha _{8}$ \\ \hline
$\sigma ^{3}$ & $\alpha _{4}+\alpha _{5}+\alpha _{6}+2\,\alpha _{7}+\alpha
_{8}$ & {\small $\phi $}$(\tilde{\alpha}_{1}+\tilde{\alpha}_{2}+2\tilde{%
\alpha}_{3}+${\small $\phi $}$\tilde{\alpha}_{4}) $ & $%
\begin{array}{c}
\alpha _{1}+\alpha _{2}+2\,\alpha _{3}+\alpha _{4} \\ 
+\alpha _{5}+\alpha _{6}+2\,\alpha _{7}+2\,\alpha _{8}%
\end{array}
$ \\ \hline
$\sigma ^{4}$ & $%
\begin{array}{c}
\alpha _{3}+\alpha _{4}+\alpha _{5} \\ 
\multicolumn{1}{r}{+2(\alpha _{6}+\alpha _{7})+\alpha _{8}}%
\end{array}
$ & {\small $\phi $}$(\tilde{\alpha}_{1}+2\,\tilde{\alpha}_{2})+${\small $%
\phi $}$^{3}\tilde{\alpha}_{3}+${\small $\phi $}$^{2}\tilde{\alpha}_{4}$ & $%
\begin{array}{c}
\alpha _{1}+2\,\alpha _{2}+2\,\alpha _{3}+\alpha _{4} \\ 
+\alpha _{5}+2\,\alpha _{6}+3\,\alpha _{7}+2\,\alpha _{8}%
\end{array}
$ \\ \hline
$\sigma ^{5}$ & $%
\begin{array}{c}
\alpha _{1}+\alpha _{2}+\alpha _{4}+\alpha _{6} \\ 
\multicolumn{1}{r}{+2(\alpha _{3}+\alpha _{7}+\alpha _{8})}%
\end{array}
$ & $\tilde{\alpha}_{1}+${\small $\phi $}$^{2}(\tilde{\alpha}_{2}+2\tilde{%
\alpha}_{3})+${\small $\phi $}$^{3}\tilde{\alpha}_{4}$ & $%
\begin{array}{c}
\alpha _{2}+2\,\alpha _{3}+2\,\alpha _{4}+\alpha _{5} \\ 
+2\,\alpha _{6}+4\,\alpha _{7}+3\,\alpha _{8}%
\end{array}
$ \\ \hline
$\sigma ^{6}$ & $%
\begin{array}{c}
\alpha _{1}+\alpha _{4}+\alpha _{5}+\alpha _{6} \\ 
+2(\alpha _{2}+\alpha _{3}+\alpha _{7}+\alpha _{8})%
\end{array}
$ & {\small $\phi $}$^{2}(\tilde{\alpha}_{1}+2\tilde{\alpha}_{3}+${\small $%
\phi $}$\tilde{\alpha}_{4})+(2+${\small $\phi $}$)\tilde{\alpha}_{2}$ & $%
\begin{array}{c}
\alpha _{1}+\alpha _{2}+2\alpha _{3}+2\alpha _{4} \\ 
+2\alpha _{5}+3\alpha _{6}+4\alpha _{7}+3\alpha _{8}%
\end{array}
$ \\ \hline
$\sigma ^{7}$ & $%
\begin{array}{c}
\alpha _{2}+\alpha _{3}+\alpha _{4}+\alpha _{5} \\ 
+3\,\alpha _{7}+2(\alpha _{6}+\alpha _{8})%
\end{array}
$ & {\small $\phi $}$\,\tilde{\alpha}_{1}+(${\small $\phi $}$^{4}-1)\tilde{%
\alpha}_{3}+${\small $\phi $}$^{3}(\tilde{\alpha}_{2}+\tilde{\alpha}_{4})$ & 
$%
\begin{array}{c}
\alpha _{1}+2\,\alpha _{2}+3\,\alpha _{3}+2\,\alpha _{4} \\ 
+\alpha _{5}+3\,\alpha _{6}+4\,\alpha _{7}+3\,\alpha _{8}%
\end{array}
$ \\ \hline
$\sigma ^{8}$ & $%
\begin{array}{c}
\alpha _{3}+\alpha _{5}+3\,\alpha _{7} \\ 
+2(\alpha _{4}+\alpha _{6}+\alpha _{8})%
\end{array}
$ & {\small $\phi $}$\,(\tilde{\alpha}_{1}+\tilde{\alpha}_{2}+2${\small $%
\phi $}$\tilde{\alpha}_{4})+(${\small $\phi $}$^{4}-1)\,\tilde{\alpha}_{3}$
& $%
\begin{array}{c}
\alpha _{1}+2\,\alpha _{2}+3\,\alpha _{3}+2\,\alpha _{4} \\ 
+\alpha _{5}+2\,\alpha _{6}+4\,\alpha _{7}+4\,\alpha _{8}%
\end{array}
$ \\ \hline
$\sigma ^{9}$ & $%
\begin{array}{c}
\alpha _{1}+\alpha _{2}+\alpha _{4}+\alpha _{5} \\ 
+2(\alpha _{3}+\alpha _{6}+\alpha _{7}+\alpha _{8})%
\end{array}
$ & {\small $\phi $}$^{2}\left( \tilde{\alpha}{_{1}}+2\tilde{\alpha}{_{3}}%
\right) +${\small $\phi $}$^{3}\,(\tilde{\alpha}_{2}+\,\tilde{\alpha}_{4})$
& $%
\begin{array}{c}
\alpha _{1}+2\,\alpha _{2}+2\,\alpha _{3}+2\,\alpha _{4} \\ 
+2\,\alpha _{5}+3\,\alpha _{6}+4\,\alpha _{7}+3\,\alpha _{8}%
\end{array}
$ \\ \hline
$\sigma ^{10}$ & $%
\begin{array}{c}
\alpha _{1}+2(\alpha _{2}+\alpha _{3})+\alpha _{4} \\ 
+\alpha _{6}+2(\alpha _{7}+\alpha _{8})%
\end{array}
$ & $\tilde{\alpha}_{1}+\left( 2+\phi \right) \tilde{\alpha}_{2}+${\small $%
\phi $}$^{2}(2\tilde{\alpha}_{3}+${\small $\phi $}$\tilde{\alpha}_{4})$ & $%
\begin{array}{c}
\alpha _{2}+2\,\alpha _{3}+2\,\alpha _{4}+\alpha _{5} \\ 
+3\,\alpha _{6}+4\,\alpha _{7}+3\,\alpha _{8}%
\end{array}
$ \\ \hline
$\sigma ^{11}$ & $%
\begin{array}{c}
\alpha _{2}+\alpha _{3}+\alpha _{4}+\alpha _{5} \\ 
+\alpha _{6}+2\,\alpha _{7}+2\,\alpha _{8}%
\end{array}
$ & {\small $\phi $}$\,\tilde{\alpha}_{1}+${\small $\phi $}$^{2}\,\tilde{%
\alpha}_{2}+${\small $\phi $}$^{3}\left( \tilde{\alpha}{_{3}}+\tilde{\alpha}{%
_{4}}\right) $ & $%
\begin{array}{c}
\alpha _{1}+\alpha _{2}+2\,\alpha _{3}+2\,\alpha _{4} \\ 
+\alpha _{5}+2\,\alpha _{6}+3\,\alpha _{7}+3\,\alpha _{8}%
\end{array}
$ \\ \hline
$\sigma ^{12}$ & $\alpha _{4}+\alpha _{5}+2(\alpha _{6}+\alpha _{7})+\alpha
_{8}$ & {\small $\phi $}$\tilde{\alpha}_{1}+2\,${\small $\phi $}$\,(\tilde{%
\alpha}_{2}+\tilde{\alpha}_{3})+${\small $\phi $}$^{2}\tilde{\alpha}_{4}$ & $%
\begin{array}{c}
\alpha _{1}+2\,\alpha _{2}+2\,\alpha _{3}+\alpha _{4} \\ 
+\alpha _{5}+2\,\alpha _{6}+2\,\alpha _{7}+2\,\alpha _{8}%
\end{array}
$ \\ \hline
$\sigma ^{13}$ & $\alpha _{3}+\alpha _{4}+\alpha _{6}+\alpha _{7}+\alpha
_{8} $ & {\small $\phi $}$\tilde{\alpha}_{2}+${\small $\phi $}$^{2}\,\left( 
\tilde{\alpha}{_{3}}+\tilde{\alpha}{_{4}}\right) $ & $\alpha _{2}+\alpha
_{3}+\alpha _{4}+\alpha _{6}+2(\alpha _{7}+\alpha _{8})$ \\ \hline
$\sigma ^{14}$ & $\alpha _{1}+\alpha _{2}+\alpha _{3}+\alpha _{8}$ & $\tilde{%
\alpha}_{1}+\tilde{\alpha}_{2}+\tilde{\alpha}_{3}+${\small $\phi $}$\tilde{%
\alpha}_{4}$ & $\alpha _{4}+\alpha _{5}+\alpha _{6}+\alpha _{7}+\alpha _{8}$
\\ \hline
$\sigma ^{15}$ & $\alpha _{2}$ & $\tilde{\alpha}_{2}$ & $\alpha _{6}$ \\ 
\hline
\end{tabular}

\begin{tabular}{||l||l|l|l||}
\hline
& $\Omega _{3}$ & $\omega (\Omega _{3})=\tilde{\Omega}_{3}$ & $\Omega _{7}$
\\ \hline
$\sigma ^{0}$ & $\alpha _{3}$ & $\tilde{\alpha}_{3}$ & $\alpha _{7}$ \\ 
\hline
$\sigma ^{1}$ & $\alpha _{1}+\alpha _{2}+\alpha _{3}+\alpha _{7}+\alpha _{8}$
& $\tilde{\alpha}_{1}+\tilde{\alpha}_{2}+${\small $\phi $}$^{2}\tilde{\alpha}%
_{3}+${\small $\phi $}$\tilde{\alpha}_{4}$ & $%
\begin{array}{c}
\alpha _{3}+\alpha _{4}+\alpha _{5} \\ 
+\alpha _{6}+2\,\alpha _{7}+\alpha _{8}%
\end{array}
$ \\ \hline
$\sigma ^{2}$ & $%
\begin{array}{c}
\alpha _{2}+\alpha _{3}+\alpha _{4}+\alpha _{5} \\ 
+\alpha _{6}+2\,\alpha _{7}+\alpha _{8}%
\end{array}
$ & {\small $\phi $}$\tilde{\alpha}_{1}+${\small $\phi $}$^{2}\tilde{\alpha}%
_{2}+${\small $\phi $}$^{3}\tilde{\alpha}_{3}+${\small $\phi $}$^{2}\tilde{%
\alpha}_{4}$ & $%
\begin{array}{c}
\alpha _{1}+\alpha _{2}+\alpha _{4}+\alpha _{5} \\ 
+3\,\alpha _{7}+2(\alpha _{3}+\alpha _{6}+\alpha _{8})%
\end{array}
$ \\ \hline
$\sigma ^{3}$ & $%
\begin{array}{c}
\alpha _{3}+\alpha _{4}+\alpha _{5} \\ 
+3\,\alpha _{7}+2(\alpha _{6}+\alpha _{8})%
\end{array}
$ & {\small $\phi $}$(\tilde{\alpha}_{1}+2\tilde{\alpha}_{2})+(${\small $%
\phi $}$^{4}-1)\tilde{\alpha}_{3}+${\small $\phi $}$^{3}\tilde{\alpha}_{4}$
& $%
\begin{array}{c}
\alpha _{1}+2(\alpha _{2}+\alpha _{4}+\alpha _{6}) \\ 
+\alpha _{5}+4\alpha _{7}+3(\alpha _{3}+\alpha _{8})%
\end{array}
$ \\ \hline
$\sigma ^{4}$ & $%
\begin{array}{c}
\alpha _{1}+\alpha _{2}+2(\alpha _{3}+\alpha _{4}) \\ 
+\alpha _{5}+3\,\alpha _{7}+2(\alpha _{6}+\alpha _{8})%
\end{array}
$ & {\small $\phi $}$^{2}(\tilde{\alpha}_{1}+${\small $\phi $}$\tilde{\alpha}%
_{2}+${\small $\phi $}$^{2}\tilde{\alpha}_{3}+2\tilde{\alpha}_{4})$ & $%
\begin{array}{c}
\alpha _{1}+2(\alpha _{2}+\alpha _{4}+\alpha _{5}) \\ 
+3(\alpha _{3}+\alpha _{6})+5\,\alpha _{7}+4\,\alpha _{8}%
\end{array}
$ \\ \hline
$\sigma ^{5}$ & $%
\begin{array}{c}
\alpha _{1}+\alpha _{4}+2(\alpha _{2}+\alpha _{6}) \\ 
+\alpha _{5}+3(\alpha _{3}+\alpha _{7}+\alpha _{8})%
\end{array}
$ & $%
\begin{array}{c}
\phi ^{2}\left( \tilde{\alpha}{_{1}}+2\,\tilde{\alpha}{_{2}}+3\,\tilde{\alpha%
}{_{3}}\right) \\ 
+\left( \phi ^{4}-1\right) \tilde{\alpha}_{4}%
\end{array}
$ & $%
\begin{array}{c}
\alpha _{1}+2\,\alpha _{2}+3(\alpha _{3}+\alpha _{4}) \\ 
+2\,\alpha _{5}+6\,\alpha _{7}+4(\alpha _{6}+\alpha _{8})%
\end{array}
$ \\ \hline
$\sigma ^{6}$ & $%
\begin{array}{c}
2(\alpha _{2}+\alpha _{3}+\alpha _{4}+\alpha _{6}) \\ 
+\alpha _{1}+\alpha _{5}+4\,\alpha _{7}+3\,\alpha _{8}%
\end{array}
$ & {\small $\phi $}$^{2}(\tilde{\alpha}_{1}+2\tilde{\alpha}_{2}+2${\small $%
\phi $}$\tilde{\alpha}_{3}+${\small $\phi $}$^{2}\tilde{\alpha}_{4})$ & $%
\begin{array}{c}
\alpha _{1}+2(\alpha _{2}+\alpha _{5})+3\,\alpha _{4} \\ 
+4(\alpha _{3}+\alpha _{6})+6\,\alpha _{7}+5\,\alpha _{8}%
\end{array}
$ \\ \hline
$\sigma ^{7}$ & $%
\begin{array}{c}
\alpha _{2}+2(\alpha _{3}+\alpha _{4}+\alpha _{5}) \\ 
+4\,\alpha _{7}+3(\alpha _{6}+\alpha _{8})%
\end{array}
$ & $%
\begin{array}{c}
\phi (\tilde{\alpha}_{1}+2\phi ^{2}\tilde{\alpha}_{3}+\phi ^{3}\tilde{\alpha}%
_{4}) \\ 
+\left( \phi ^{4}-1\right) \tilde{\alpha}_{2}%
\end{array}
$ & $%
\begin{array}{c}
2(\alpha _{1}+\alpha _{5})+3(\alpha _{2}+\alpha _{4}) \\ 
+4(\alpha _{3}+\alpha _{6})+6\,\alpha _{7}+5\,\alpha _{8}%
\end{array}
$ \\ \hline
$\sigma ^{8}$ & $%
\begin{array}{c}
\alpha _{1}+\alpha _{2}+2(\alpha _{3}+\alpha _{4}) \\ 
+\alpha _{5}+4\,\alpha _{7}+3(\alpha _{6}+\alpha _{8})%
\end{array}
$ & $%
\begin{array}{c}
\phi ^{2}(\tilde{\alpha}_{1}+2\phi \tilde{\alpha}_{3}+\phi ^{2}\tilde{\alpha}%
_{4}) \\ 
+\left( \phi ^{4}-1\right) \tilde{\alpha}_{2}%
\end{array}
$ & $%
\begin{array}{c}
\alpha _{1}+3(\alpha _{2}+\alpha _{4})+2\,\alpha _{5} \\ 
+4(\alpha _{3}+\alpha _{6})+6\,\alpha _{7}+5\,\alpha _{8}%
\end{array}
$ \\ \hline
$\sigma ^{9}$ & $%
\begin{array}{c}
\alpha _{1}+2(\alpha _{2}+\alpha _{4}+\alpha _{6}) \\ 
+\alpha _{5}+3(\alpha _{3}+\alpha _{7}+\alpha _{8})%
\end{array}
$ & {\small $\phi $}$^{2}\left( \tilde{\alpha}{_{1}}+2\,\tilde{\alpha}{_{2}}%
+3\,\tilde{\alpha}{_{3}}\right) +${\small $\phi $}$^{4}\tilde{\alpha}_{4}$ & 
$%
\begin{array}{c}
\alpha _{1}+2(\alpha _{2}+\,\alpha _{5})+4\,\alpha _{6} \\ 
+3(\alpha _{3}+\alpha _{4})+6\,\alpha _{7}+5\,\alpha _{8}%
\end{array}
$ \\ \hline
$\sigma ^{10}$ & $%
\begin{array}{c}
\alpha _{1}+2(\alpha _{2}+\alpha _{3}+\alpha _{6}) \\ 
+\alpha _{4}+\alpha _{5}+3(\alpha _{7}+\alpha _{8})%
\end{array}
$ & $%
\begin{array}{c}
\phi ^{2}(\tilde{\alpha}_{1}+2\tilde{\alpha}_{2}+\phi ^{2}\tilde{\alpha}_{3})
\\ 
+\left( \phi ^{4}-1\right) \tilde{\alpha}_{4}%
\end{array}
$ & $%
\begin{array}{c}
\alpha _{1}+2\,(\alpha _{2}+\alpha _{5})+4\,\alpha _{6} \\ 
+3(\alpha _{3}+\alpha _{4})+5\,\alpha _{7}+4\,\alpha _{8}%
\end{array}
$ \\ \hline
$\sigma ^{11}$ & $%
\begin{array}{c}
\alpha _{2}+\alpha _{3}+\alpha _{5}+3\alpha _{7} \\ 
+2(\alpha _{4}+\alpha _{6}+\alpha _{8})%
\end{array}
$ & {\small $\phi $}$\,\tilde{\alpha}_{1}+${\small $\phi $}$^{3}\tilde{\alpha%
}_{2}+(${\small $\phi $}$^{4}-1)\tilde{\alpha}_{3}+2${\small $\phi $}$^{2}%
\tilde{\alpha}_{4}$ & $%
\begin{array}{c}
\alpha _{1}+2(\alpha _{2}+\alpha _{4})+\alpha _{5} \\ 
+3(\alpha _{3}+\alpha _{6})+4(\alpha _{7}+\alpha _{8})%
\end{array}
$ \\ \hline
$\sigma ^{12}$ & $%
\begin{array}{c}
\alpha _{3}+\alpha _{4}+\alpha _{5} \\ 
+2(\alpha _{6}+\alpha _{7}+\alpha _{8})%
\end{array}
$ & {\small $\phi $}$\tilde{\alpha}_{1}+2${\small $\phi $}$\tilde{\alpha}%
_{2}+${\small $\phi $}$^{3}\left( \tilde{\alpha}{_{3}}+\tilde{\alpha}{_{4}}%
\right) $ & $%
\begin{array}{c}
\alpha _{1}+2(\alpha _{2}+\alpha _{3}+\alpha _{4}) \\ 
+\alpha _{5}+2\,\alpha _{6}+3(\alpha _{7}+\alpha _{8})%
\end{array}
$ \\ \hline
$\sigma ^{13}$ & $%
\begin{array}{c}
\alpha _{1}+\alpha _{2}+\alpha _{3}+\alpha _{4} \\ 
+\alpha _{6}+\alpha _{7}+\alpha _{8}%
\end{array}
$ & $\tilde{\alpha}_{1}+${\small $\phi $}$^{2}\left( \tilde{\alpha}{_{2}}+%
\tilde{\alpha}{_{3}}+\tilde{\alpha}{_{4}}\right) $ & $%
\begin{array}{c}
\alpha _{2}+\alpha _{3}+\alpha _{4}+\alpha _{5} \\ 
+2(\alpha _{6}+\alpha _{7}+\alpha _{8})%
\end{array}
$ \\ \hline
$\sigma ^{14}$ & $\alpha _{2}+\alpha _{3}+\alpha _{8}$ & $\tilde{\alpha}_{2}+%
\tilde{\alpha}_{3}+${\small $\phi $}$\tilde{\alpha}_{4}$ & $\alpha
_{4}+\alpha _{6}+\alpha _{7}+\alpha _{8}$ \\ \hline
$\sigma ^{15}$ & $-\alpha _{3}$ & $-\tilde{\alpha}_{3}$ & $-\alpha _{7}$ \\ 
\hline
\end{tabular}

\begin{tabular}{||l||l|l|l||}
\hline
& $\Omega _{4}$ & $\omega (\Omega _{4})=\tilde{\Omega}_{4}$ & $\Omega _{8}$
\\ \hline
$\sigma ^{0}$ & $-\alpha _{4}$ & $-\tilde{\alpha}_{4}$ & $-\alpha _{8}$ \\ 
\hline
$\sigma ^{1}$ & $\alpha _{4}+\alpha _{7}$ & {\small $\phi $}$\tilde{\alpha}%
_{3}+\tilde{\alpha}_{4}$ & $\alpha _{3}+\alpha _{7}+\alpha _{8}$ \\ \hline
$\sigma ^{2}$ & $\alpha _{3}+\alpha _{5}+\alpha _{6}+\alpha _{7}+\alpha _{8}$
& {\small $\phi $}$\left( \tilde{\alpha}{_{1}}+\tilde{\alpha}{_{2}}+\phi 
\tilde{\alpha}{_{3}}+\tilde{\alpha}{_{4}}\right) $ & $%
\begin{array}{c}
\alpha _{1}+\alpha _{2}+\alpha _{3}+\alpha _{4} \\ 
+\alpha _{5}+\alpha _{6}+2\,\alpha _{7}+\alpha _{8}%
\end{array}
$ \\ \hline
$\sigma ^{3}$ & $%
\begin{array}{c}
\alpha _{1}+\alpha _{2}+\alpha _{3}+\alpha _{4} \\ 
+\alpha _{6}+2\,\alpha _{7}+\alpha _{8}%
\end{array}
$ & $\tilde{\alpha}_{1}+${\small $\phi $}$^{2}(\tilde{\alpha}_{2}+\tilde{%
\alpha}_{4})+${\small $\phi $}$^{3}\tilde{\alpha}_{3}$ & $%
\begin{array}{c}
\alpha _{2}+\alpha _{4}+\alpha _{5}+3\,\alpha _{7} \\ 
+2(\alpha _{3}+\alpha _{6}+\alpha _{8})%
\end{array}
$ \\ \hline
$\sigma ^{4}$ & $%
\begin{array}{c}
\alpha _{2}+\alpha _{4}+\alpha _{5}+\alpha _{6} \\ 
+2(\alpha _{3}+\alpha _{7}+\alpha _{8})%
\end{array}
$ & {\small $\phi $}$\tilde{\alpha}_{1}+${\small $\phi $}$^{2}\left( \tilde{%
\alpha}{_{2}}+2\,\tilde{\alpha}{_{3}+\phi }\tilde{\alpha}{_{4}}\right) $ & $%
\begin{array}{c}
\alpha _{1}+\alpha _{2}+2(\alpha _{3}+\alpha _{4}+\alpha _{6}) \\ 
+\alpha _{5}+4\,\alpha _{7}+3\,\alpha _{8}%
\end{array}
$ \\ \hline
$\sigma ^{5}$ & $%
\begin{array}{c}
\alpha _{1}+\alpha _{2}+\alpha _{3}+\alpha _{4} \\ 
+\alpha _{5}+3\alpha _{7}+2(\alpha _{6}+\alpha _{8})%
\end{array}
$ & $%
\begin{array}{c}
\phi ^{2}(\tilde{\alpha}_{1}+\phi \,\tilde{\alpha}_{2}+\phi \tilde{\alpha}%
_{4}) \\ 
+(\phi ^{4}-1)\tilde{\alpha}_{3}%
\end{array}
$ & $%
\begin{array}{c}
\alpha _{1}+2(\alpha _{2}+\alpha _{4}+\alpha _{5}) \\ 
+3(\alpha _{3}+\alpha _{6}+\alpha _{8})+4\,\alpha _{7}%
\end{array}
$ \\ \hline
$\sigma ^{6}$ & $%
\begin{array}{c}
\alpha _{2}+2(\alpha _{3}+\alpha _{4})+\alpha _{5} \\ 
+3\,\alpha _{7}+2(\alpha _{6}+\alpha _{8})%
\end{array}
$ & {\small $\phi $}$(\tilde{\alpha}_{1}+${\small $\phi $}$^{2}\tilde{\alpha}%
_{2}+${\small $\phi $}$^{3}\tilde{\alpha}_{3}+${\small $\phi $}$\tilde{\alpha%
}_{4})$ & $%
\begin{array}{c}
\alpha _{1}+2(\alpha _{2}+\alpha _{4})+\alpha _{5} \\ 
+3(\alpha _{3}+\alpha _{6})+5\,\alpha _{7}+4\,\alpha _{8}%
\end{array}
$ \\ \hline
$\sigma ^{7}$ & $%
\begin{array}{c}
\alpha _{1}+\alpha _{2}+2(\alpha _{3}+\alpha _{6}) \\ 
+\alpha _{4}+\alpha _{5}+3(\alpha _{7}+\alpha _{8})%
\end{array}
$ & $%
\begin{array}{c}
\phi ^{2}(\tilde{\alpha}_{1}+\phi \tilde{\alpha}_{2}+\phi ^{2}\tilde{\alpha}%
_{3}) \\ 
+(\phi ^{4}-1)\tilde{\alpha}_{4}%
\end{array}
$ & $%
\begin{array}{c}
\alpha _{1}+2(\alpha _{2}+\alpha _{5})+5\alpha _{7} \\ 
+3(\alpha _{3}+\alpha _{4}+\alpha _{6})+4\,\alpha _{8}%
\end{array}
$ \\ \hline
$\sigma ^{8}$ & $%
\begin{array}{c}
\alpha _{1}+2(\alpha _{2}+\alpha _{3}+\alpha _{4}) \\ 
+\alpha _{5}+2(\alpha _{6}+\alpha _{8})+3\alpha _{7}%
\end{array}
$ & {\small $\phi $}$^{2}(\tilde{\alpha}_{1}+2\tilde{\alpha}_{2}+${\small $%
\phi $}$^{2}\tilde{\alpha}_{3}+2\tilde{\alpha}_{4})$ & $%
\begin{array}{c}
\alpha _{1}+2(\alpha _{2}+\alpha _{4}+\alpha _{5}) \\ 
+3\,\alpha _{3}+5\,\alpha _{7}+4(\alpha _{6}+\alpha _{8})%
\end{array}
$ \\ \hline
$\sigma ^{9}$ & $%
\begin{array}{c}
\alpha _{2}+\alpha _{4}+\alpha _{5} \\ 
+2(\alpha _{6}+\alpha _{3})+3(\alpha _{7}+\alpha _{8})%
\end{array}
$ & $%
\begin{array}{c}
\phi \tilde{\alpha}_{1}+\phi ^{3}\tilde{\alpha}_{2}+\phi ^{4}\tilde{\alpha}%
_{3} \\ 
+(\phi ^{4}-1)\tilde{\alpha}_{4}%
\end{array}
$ & $%
\begin{array}{c}
\alpha _{1}+2\,\alpha _{2}+\alpha _{5}+5\alpha _{7} \\ 
+3(\alpha _{3}+\alpha _{4}+\alpha _{6})+4\,\alpha _{8}%
\end{array}
$ \\ \hline
$\sigma ^{10}$ & $%
\begin{array}{c}
\alpha _{1}+\alpha _{2}+\alpha _{3}+\alpha _{5} \\ 
+2(\alpha _{4}+\alpha _{6}+\alpha _{8})+3\alpha _{7}%
\end{array}
$ & $%
\begin{array}{c}
\phi ^{2}(\tilde{\alpha}_{1}+2\tilde{\alpha}_{4})+\phi ^{3}\tilde{\alpha}_{2}
\\ 
+\left( \phi ^{4}-1\right) \tilde{\alpha}_{3}%
\end{array}
$ & $%
\begin{array}{c}
\alpha _{1}+2(\alpha _{2}+\alpha _{4}+\alpha _{5}) \\ 
+3(\alpha _{3}+\alpha _{6})+4(\alpha _{7}+\alpha _{8})%
\end{array}
$ \\ \hline
$\sigma ^{11}$ & $%
\begin{array}{c}
\alpha _{2}+\alpha _{4}+\alpha _{5} \\ 
+2(\alpha _{3}+\alpha _{6}+\alpha _{7}+\alpha _{8})%
\end{array}
$ & $%
\begin{array}{c}
\phi \,\tilde{\alpha}_{1}+\phi ^{3}(\tilde{\alpha}_{2}+\tilde{\alpha}_{4})
\\ 
+2\phi ^{2}\,\tilde{\alpha}_{3}%
\end{array}
$ & $%
\begin{array}{c}
\alpha _{1}+2(\alpha _{2}+\alpha _{3}+\alpha _{4}) \\ 
+\alpha _{5}+4\,\alpha _{7}+3(\alpha _{6}+\alpha _{8})%
\end{array}
$ \\ \hline
$\sigma ^{12}$ & $%
\begin{array}{c}
\alpha _{1}+\alpha _{2}+\alpha _{3}+\alpha _{4} \\ 
+\alpha _{6}+2(\alpha _{7}+\alpha _{8})%
\end{array}
$ & $\tilde{\alpha}_{1}+${\small $\phi $}$^{2}\,\tilde{\alpha}_{2}+${\small $%
\phi $}$^{3}\,\left( \tilde{\alpha}{_{3}}+\tilde{\alpha}{_{4}}\right) $ & $%
\begin{array}{c}
\alpha _{2}+2(\alpha _{3}+\alpha _{4}+\alpha _{6}) \\ 
+\alpha _{5}+3(\alpha _{7}+\alpha _{8})%
\end{array}
$ \\ \hline
$\sigma ^{13}$ & $%
\begin{array}{c}
\alpha _{2}+\alpha _{3}+\alpha _{4}+\alpha _{5} \\ 
+\alpha _{6}+\alpha _{7}+\alpha _{8}%
\end{array}
$ & {\small $\phi $}$\,\tilde{\alpha}_{1}+${\small $\phi $}$^{2}\left( 
\tilde{\alpha}{_{2}}+\tilde{\alpha}{_{3}}+\tilde{\alpha}{_{4}}\right) $ & $%
\begin{array}{c}
\alpha _{1}+\alpha _{2}+\alpha _{3}+\alpha _{4} \\ 
+\alpha _{5}+2(\alpha _{6}+\alpha _{7}+\alpha _{8})%
\end{array}
$ \\ \hline
$\sigma ^{14}$ & $\alpha _{6}+\alpha _{7}+\alpha _{8}$ & {\small $\phi $}$%
\left( \tilde{\alpha}{_{2}}+\tilde{\alpha}{_{3}}+\tilde{\alpha}{_{4}}\right) 
$ & $\alpha _{2}+\alpha _{3}+\alpha _{4}+\alpha _{6}+\alpha _{7}+\alpha _{8}$
\\ \hline
$\sigma ^{15}$ & $\alpha _{4}$ & $\tilde{\alpha}_{4}$ & $\alpha _{8}$ \\ 
\hline
\end{tabular}
\newpage
\end{center}

\bibliographystyle{phreport}
\bibliography{Ref}

\end{document}